\documentclass[aps,physrev,reprint,showpacs,longbibliography]{revtex4-2}
\usepackage[english]{babel}
\usepackage{amsmath,amssymb,bbm,graphicx,color,comment}
\usepackage[bookmarks=true,colorlinks,citecolor=blue,urlcolor=blue]{hyperref}
\usepackage{dsfont}
\usepackage{soul}
\usepackage{bbm}
\usepackage{float}
\usepackage{dcolumn}   % needed for some tables
\usepackage{bm}        % for math
\usepackage{filecontents}
\usepackage{lineno}
\usepackage{mathtools}
\usepackage[usenames,dvipsnames]{xcolor} 
\usepackage[export]{adjustbox}
\usepackage{adjustbox}
\usepackage{braket}
\usepackage{tikz}
\usepackage{bbold}

\usepackage{tikz}
\usetikzlibrary{fit}
\usetikzlibrary{matrix}

 \renewcommand{\v}[1]{\ensuremath{\mathbf{#1}}}

\newcommand{\bs}[1]{\boldsymbol{#1}}

\let\baraccent=\= \renewcommand{\=}[1]{\stackrel{#1}{=}}
\newcommand{\eq}[1]{Eq.\thinspace(\ref{#1})}
\newcommand{\eqs}[2]{Eqs.\thinspace(\ref{#1},\ref{#2})}

\newcommand{\fig}[1]{Fig.\thinspace{}\ref{#1}}
\newcommand{\fc}[1]{({#1})}
\newcommand{\figc}[2]{Fig.\thinspace{}\ref{#1}\thinspace{}\fc{#2}}

\usepackage[normalem]{ulem}

\begin{document}
\title{
Quantum simulation of dynamical gauge theories in periodically \\ driven Rydberg atom arrays
}

\author{Johannes Feldmeier$^{1}$}
\author{Nishad Maskara$^{1}$}
\author{Nazl\i \ U\u{g}ur K\"oyl\"uo\u{g}lu$^{1,2}$}
\author{Mikhail D. Lukin$^{1}$}
\affiliation{$^1$Physics Department, Harvard University, Oxford St. 17, 02138 Cambridge MA, USA \\
$^2$Harvard Quantum Initiative, Harvard University, Cambridge, MA 02138, USA}

\date{\today}

\begin{abstract}
Simulating quantum dynamics of lattice gauge theories (LGTs) is an exciting frontier in quantum science. Programmable quantum simulators based on neutral atom arrays are a promising approach to achieve this goal, since strong Rydberg blockade interactions can be used to naturally create low energy subspaces that can encode local gauge constraints. However, realizing 
regimes of LGTs where both matter and gauge fields exhibit significant dynamics requires the presence of tunable multi-body interactions such as those associated with ring exchange, which are challenging to realize directly.  
Here, we develop a method for generating such interactions based on time-periodic driving. Our approach utilizes controlled deviations from time-reversed trajectories, which are accessible in constrained PXP-type models via the application of frequency modulated global pulses. We show that such driving gives rise to a family of effective Hamiltonians with multi-body interactions whose strength is non-perturbative in their respective operator weight.
We apply this approach to a two-dimensional U(1) LGT on the Kagome lattice, where we engineer strong six-body magnetic plaquette terms that are tunable relative to the kinetic energy of matter excitations, demonstrating access to previously unexplored dynamical regimes. Potential generalizations and prospects for experimental implementations are discussed. 
\end{abstract}

%\pacs{Valid PACS appear here}
\maketitle
%\tableofcontents

\section{Introduction}

\begin{figure}[t]
\centering
\includegraphics[trim={0cm 0cm 0cm 0cm},clip,width=0.99\linewidth]{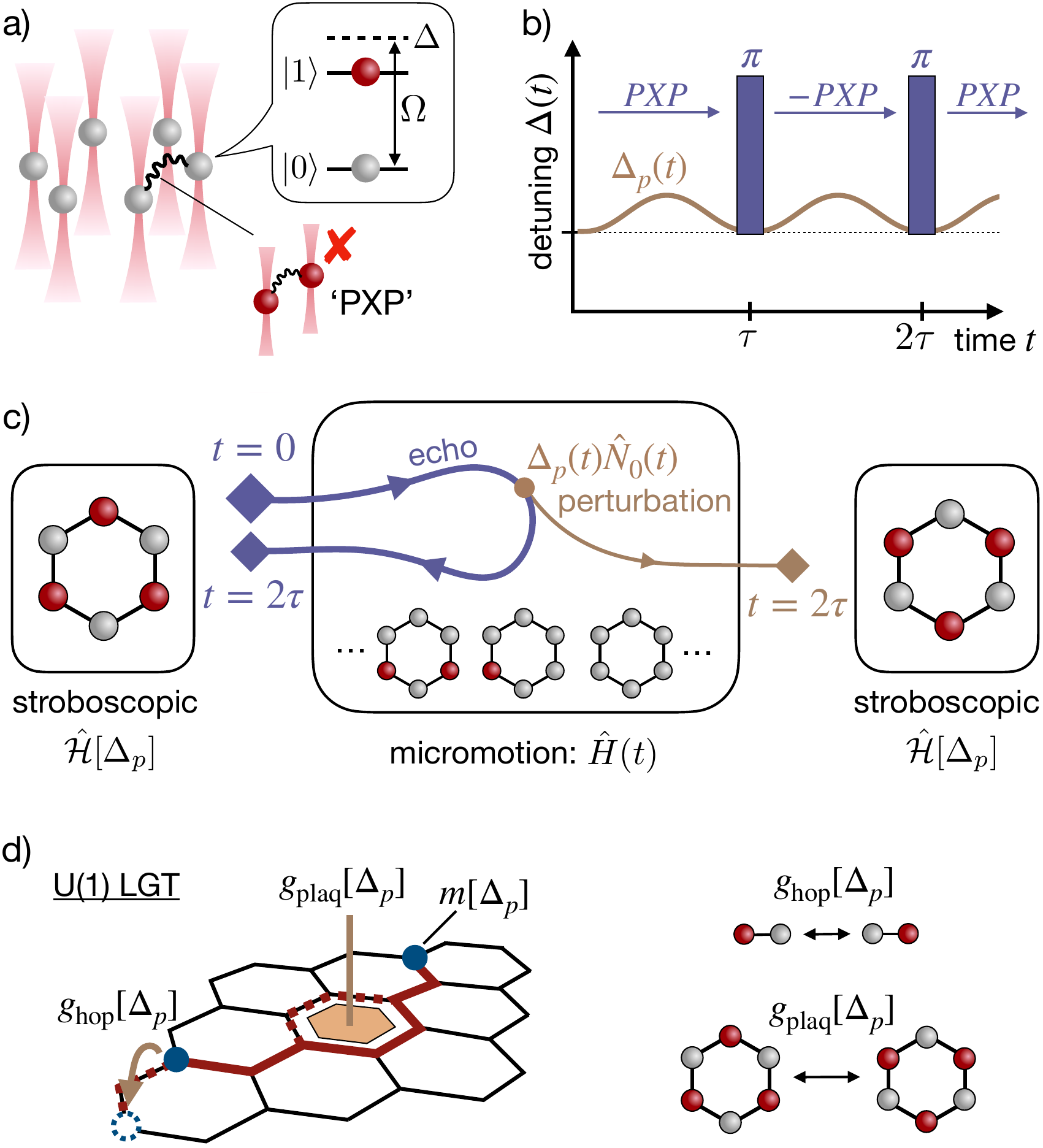}
\caption{\textbf{Hamiltonian engineering and gauge theories.} \textbf{a)} We consider PXP models as an approximate description of Rydberg atoms in optical tweezers. \textbf{b)} Due to blockade, simple global detuning $\pi$-pulses reverse the direction of time evolution, leading to an effective many-body echo. \textbf{c)} Introducing detuning perturbations $\Delta_p(t)$ around this echo realizes an effective Hamiltonian that consists of the time-evolved Rydberg number operators $\hat{N}_0(t)$. The spreading of this operator leads to multi-body interactions in the effective Hamiltonian. Intuitively, for an optimized choice of $\Delta_p(t)$, part of the wave function `scatters' into a desired final state, realizing a coherent interaction term. \textbf{d)} Our approach can be applied to study lattice gauge theories, where multi-body plaquette terms are required to realize dynamical gauge fields. We achieve increased flexibility in tuning the relative strength of various interactions, enabling access to regimes previously difficult to realize.
}
\label{fig:1}
\end{figure}

A central challenge in the field of quantum simulations is the programmable realization of a wide range of many-body systems on current devices. 
Within any given approach, the primary control tools for engineering specific quantum many-body Hamiltonians are typically provided by geometric configuration and time-dependent classical control of external fields.  
Neutral atoms in optical tweezers constitute a promising platform for programmable quantum simulations with an exceptional degree of geometric flexibility~\cite{Bernien2017_51atom,Browaeys2020_rydberg,Labuhn2016_rydberg,morgado2021quantum}. In combination with Rydberg blockade interactions, highly constrained Hilbert spaces can be realized as emergent low energy subspaces. These can in turn give rise to exotic phases of matter including spin liquid states~\cite{Bernien2017_51atom,ebadi2021_256atom,semeghini2021_sl}, novel out-of-equilibrium phenomena such as quantum many-body scars~\cite{Bernien2017_51atom,bluvstein2021_scars,turner2018}, and can be used to encode classical optimization problems~\cite{ebad2022_mis,pichler2018quantum} or local gauge constraints~\cite{glaetzle2014_ice,surace2020_lgt,celi2020_lgt,verresen2021_sl,samajdar2023_z2,Homeier_2023vt,zeng2024quantum}. The latter open the door to probing the nonequilibrium dynamics of lattice gauge theories (LGTs) beyond one dimension, an important quantum simulation goal with relevance to high energy physics and emergent phenomena in condensed matter~\cite{wiese2013ultracold,dalmonte2016lattice,zohar2015quantum,
banuls2020simulating,aidelsburger2022cold}.

However, a remaining key challenge for realizing these models is the generation of interesting dynamics while simultaneously preserving the local constraints. In particular, energetic conditions that stabilize a desired subspace may also lead to strong suppression of subspace-preserving evolution, which typically arises through high-order, multi-body perturbative processes. 
Consequently, the limited tunability of perturbation theory poses an inherent difficulty to simulating different regimes of dynamical gauge field theories. 
A specific manifestation of this challenge is the implementation of magnetic plaquette terms with analog quantum simulators~\cite{Dai_2017ts}, corresponding to multi-body interactions in the electric field representation. While many advances in the quantum simulation of LGTs have been achieved across a variety of platforms in recent years~\cite{Bernien2017_51atom,Martinez2016_lgt,kokail2019_schwinger,
wang2022_gauge,mildenberger2022probing,Schweizer2019_lgt,
mil2020_lgt,Yang2020_gauge,Frolian2022_topo}, a direct, large-scale implementation of such interactions remains difficult.

In this work, we show that combining Rydberg blockade and periodic driving allows for the realization of dynamical gauge field theories based on tunable multi-body interactions. The central idea is related to the observation of Ref.~\cite{bluvstein2021_scars} that periodic driving can act as a many-body echo and tuning knob for stabilizing the dynamics of the so-called quantum many-body scars in the `PXP' models associated with the Rydberg blockade.  These observations indicate the possibility that Floquet-engineering~\cite{goldman2014_floquet,bukov2015universal,eckardt2017_floquet,
aidelsburger2013_hofstadter,miyake2013_harper,Jotzu2014ug,
flaeschner2016_floquet,meinert2016_floquet,
geier2021_floquet,scholl2022_floquet} can provide a potent tool for controlling many-body dynamics in systems with Rydberg blockade~\cite{maskara2021_dtc,hudomal2022_drive,ljubotina2022_steering}.
Building on these insights, we develop a
%More specifically, our 
method involving the use of a Floquet perturbation theory around closed periodic trajectories generated by a many-body echo. Within this framework, initial states periodically revive under continued forward- and backward-evolution generated by a blockade-consistent PXP model. The effect of deviations from an exact many-body echo are described by an effective Hamiltonian~\cite{maskara2021_dtc} with multi-body interactions, which can be \textit{engineered} from local operators dressed by interacting time evolution, see \fig{fig:1}.
Making use of operator spreading within a finite time window, multi-body interactions are generated and can be controlled via the choice of a global detuning profile. 
Small operator evolution times allow for a perturbative expansion of the resulting Hamiltonian, an approach which we use in Ref.~\cite{koyluoglu2024_ent} to implement novel, blockade-consistent spin exchange interactions. 

In this work, we develop and leverage a numerical optimization technique that enables an extension of this Floquet engineering protocol to intermediate time scales.
Crucially, this scheme takes into account the role of the detuning profile as a perturbation around the echo evolution, which we balance with the practical requirement of a substantial prefactor for the effective Hamiltonian.
We find that interactions involving a moderate number of spins are generated non-perturbatively, enabling a hardware efficient realization of many-body systems in previously inaccessible regimes. 

We subsequently apply our approach to the implementation of a two-dimensional $U(1)$ lattice gauge theory containing dynamical gauge and matter degrees of freedom.
The local gauge constraints are realized through nearest-neighbor blockade on a Kagome lattice, and we use our Floquet protocol to engineer the strengths of both gauge and matter interactions in this setup.
Since gauge dynamics are the most challenging, we further present two optimized schemes for engineering the six-body terms required to generate gauge field dynamics. In particular, we are able to access the most interesting strong coupling regime in which gauge and matter field dynamics are of comparable strength and we numerically explore their dynamical interplay.
We conclude  by discussing the realization of our PXP-based approach in arrays of Rydberg atoms with long range van-der-Waals interactions and realistic constraints on the available pulse profiles. We show how these practical challenges can be addressed and outline promising directions for future work.

Finally, before proceeding with the remainder of this work, we want to emphasize that related pulse control techniques are widely used for decoupling interactions in NMR and for Hamiltonian engineering e.g. in dipolar-interacting spin ensembles~\cite{wahuha1968,wei2018_nmr,choi2020_pulses,zhou2020_metro,
geier2024time}. In these settings, the effective Hamiltonian is obtained by averaging over site-local basis rotations. Here, we instead average over rotations generated by an interacting evolution, which enables control of larger weight operators. 
In addition, the stroboscopic implementation of a finite time step using an optimized detuning profile is reminiscent of pulse control techniques for applying two- or multi-qubit gates in digital neutral atom quantum circuits~\cite{levine2019parallel,jandura2022time,pagano2022_budget,
evered2023_gates,ma2023_gates,kalinowski2023_sl,
maskara2023_chemistry}.
Indeed, our protocol is a way of generalizing this approach to the many-body limit, while avoiding direct optimization of an exponentially large many-body unitary by restricting to finite evolution times.

\section{Floquet protocol and effective Hamiltonian} \label{sec:method}
 
To illustrate the key idea we consider a time-dependent PXP model on an arbitrary lattice geometry,
\begin{equation} \label{eq:2.0.1}
\hat{H}(t) = \frac{\Omega}{2}\sum_i \hat{P}\hat{\sigma}^x_i\hat{P} - \Delta(t) \sum_i \hat{n}_i,
\end{equation}
where $n_i\in \{0,1\}$ labels the occupation number at site $i$, $\Omega$ is the Rabi frequency and $\Delta(t)$ a time-dependent detuning coupling to the global occupation number $\hat{N}=\sum_i \hat{n}_i$. The operator $\hat{P}$ projects onto the space of configurations without nearest neighbor sites simultaneously occupying the $n_i=1$ state; explicitly, $\hat{P}=\prod_{\braket{i,j}} (1-\hat{n}_i \hat{n}_j)$, where the product runs over nearest neighbor sites $\braket{i,j}$ of the lattice. 
\eq{eq:2.0.1} is commonly used as an approximation for the dynamics of Rydberg atoms subject to strong van der Waals interactions that decay rapidly as $V_{ij}=\Omega (R_b/r_{ij})^6$: Two atoms at a distance $r_{ij}$ smaller than the blockade radius $R_b$ are effectively forbidden to both occupy the Rydberg state, thus satisfying the constraint enforced by the projector $\hat{P}$. 

\subsection{Many-body echo and effective Hamiltonian} \label{sec:echo}
The starting point for our construction is the ability to generate a closed periodic trajectory through time-dependent control of the detuning operator.
Specifically, we can reverse the sign of the off-diagonal (PXP) part of \eq{eq:2.0.1},
\begin{equation} \label{eq:pxp_model}
\hat{H}_{0}=\frac{\Omega}{2}\sum_i \hat{P}\hat{\sigma}^x_i\hat{P},
\end{equation}
by applying a global detuning pulse operator $e^{i\pi \hat{N}} = \prod_i \hat{\sigma}^z_i$, such that $e^{-i\pi\hat{N}}\hat{H}_0 \,e^{i\pi\hat{N}} = -\hat{H}_0$.
This pulse thus generates a many-body echo and repeated application at regular spacing $\tau$ leads to periodic revivals with period $2\tau$.
We denote the time-dependent Hamiltonian corresponding to this many-body echo protocol as
\begin{equation} \label{eq:2.1.0}
\hat{H}_e(t) = \hat{H}_0 - \pi \hat{N} \sum_{m \in \mathbb{N}} \delta(t-m\tau).
\end{equation}
The unitary evolution operator $\hat{U}_e(t)=\hat{\mathcal{T}}\exp\bigl\{-i\int_0^t dt^\prime \hat{H}_e(t)\bigr\}$ then indeed reduces to the identity at multiples of $2\tau$:
\begin{equation} \label{eq:2.1.1}
\begin{split}
\hat{U}_e(2\tau) = e^{-i\pi\hat{N}}e^{-i\hat{H}_{0}\tau}e^{-i\pi\hat{N}}e^{-i\hat{H}_{0}\tau} = \hat{\mathbb{1}},
\end{split}
\end{equation}
where we assume the pulses to be applied infinitesimally prior to the times $m\tau$. We emphasize the special role of the blockade in this construction: The model \eq{eq:pxp_model} is a strongly interacting, non-integrable system, yet can be dynamically decoupled using a simple $\pi$-pulse sequence, as would more commonly be used for \textit{non-interacting} disorder fields.

The stroboscopic time evolution at multiples of $2\tau$ is described by an effective Hamiltonian, $\hat{U}_e(2\tau m) = e^{-i2\tau m \hat{\mathcal{H}}}$. 
Since \eq{eq:2.1.0} generates a perfect many-body echo, $\hat{\mathcal{H}} = 0$ is trivial.
However, by adding perturbations to \eq{eq:2.1.0}, we can controllably generate a variety of terms in the effective Hamiltonian. 
In order to show this, we consider a modified detuning profile,
\begin{equation} \label{eq:2.2.1}
\Delta(t) = \pi \sum_{m \in \mathbb{N}} \delta(t-m\tau) + \Delta_p(t),
\end{equation}
where $\Delta_p(t+2\tau)=\Delta_p(t)$ denotes a $2\tau$-periodic deviation from the pure echo. 
Moreover, we restrict to profiles $\Delta_p(t)=\Delta_p(2\tau-t)$ \textit{symmetric} around $t=\tau$, as contributions to $\Delta_p$ antisymmetric around $\tau$ can always be absorbed into a redefinition of the echo evolution $\hat{U}_e(t)$.
Following Refs.~\cite{else2017_dtc,maskara2021_dtc}, we move to an interaction picture with respect to the time evolution $\hat{U}_e(t)$ and perform a high-frequency expansion~\cite{bukov2015universal}, see Appendix~\ref{sec:app_Heff} for details. Then, to leading order in the perturbation $\Delta_p(t)$, the effective Hamiltonian governing the stroboscopic time evolution is given by
\begin{equation} \label{eq:2.2.2}
\hat{\mathcal{H}}[\Delta_p] = -\int_{0}^{\tau} \frac{dt}{\tau} \Delta_p(t) \hat{N}_0(t).
\end{equation}
Here, $\hat{N}_0(t) = e^{i\hat{H}_0 t}\, \hat{N} \, e^{-i\hat{H}_0 t}$ is the global number operator time-evolved under $\hat{H}_0$. 
Using results from the theory of prethermalization~\cite{abanin2015_slow,kuwahara2016_driven,takashi2016_bound,
Abanin2017_rigor,abanin2017_effect}, in particular Ref.~\cite{else2017_dtc}, $\hat{\mathcal{H}}$ is guaranteed to be the leading order expression for an approximate, static Hamiltonian description of the time evolution on a prethermal time scale $t_*\gtrsim \exp\{C/(\sqrt{\tau}\|\Delta_p\|)\}$, with $\|\Delta_p\| \equiv \bigl(\int_0^{\tau}dt\,|\Delta_p(t)|^2 \bigr)^{1/2}$ the two-norm of $\Delta_p$ over one period of the drive and a constant $C>0$, see Appendix~\ref{sec:app_Heff}. Beyond this time scale the system is expected to eventually thermalize to infinite temperature. Here, our goal is to control the dynamics for times smaller than $t_*$ via its dominant contribution of \eq{eq:2.2.2}. 

\subsection{Optimizing the detuning profile} \label{sec:opt1}
The effective Hamiltonian $\hat{\mathcal{H}}$ in \eq{eq:2.2.2} is a linear combination of the time-evolved number operators $\hat{N}_0(t)$. Under this evolution, the operator weight of $\hat{N}_0(t)$ grows, producing multi-body interactions in $\hat{\mathcal{H}}$. In the following, we develop a formalism for controlling such terms through selection of suitable detuning profiles $\Delta_p(t)$. 
Specifically, our goal is to optimize the choice of $\Delta_p(t)$ such that $\hat{\mathcal{H}}$ is as close as possible to a desired \textit{target Hamiltonian} $\hat{T}$. 
Their distance can be quantified most directly by the standard (Frobenius) two-norm $\| \hat{T} - \hat{\mathcal{H}}[\Delta_p] \|$, taking into account all matrix elements between states in Hilbert space. 
However, in practice, we often optimize $\hat{\mathcal{H}}$ only with respect to a \textit{subset} $S$ of all such matrix elements. Accordingly, the above two-norm will be restricted to matrix elements in $S$, see Appendix~\ref{sec:optimization}. 
For example, we may be interested in optimizing matrix elements of $\hat{\mathcal{H}}$ between low-energy states of a Hamiltonian we attempt to engineer. 
Further, reducing the number of matrix elements renders the optimization of $\hat{\mathcal{H}}[\Delta_p]$ more tractable.

Then, we optimize the detuning profile by minimizing the cost function
\begin{equation} \label{eq:2.3.2}
C_{\lambda,\tau}(\hat{T}) \equiv \bigl\| \hat{T} - \hat{\mathcal{H}}[\Delta_p] \bigr\|^2 + \lambda \,\bigl\|\Delta_p\bigr\|^2.
\end{equation}
This cost function has two \textit{hyperparameters}, the Floquet period $\tau$ and a regularization parameter $\lambda$. The purpose of regularization is to ensure the norm of the perturbation remains small, while simultaneously attempting to achieve the best approximation to the target Hamiltonian $\hat{T}$~\footnote{We define the energies in \eq{eq:2.3.2} to be given in units of the Rabi frequency $\Omega$.}.
In particular, large values of $\|\Delta_p\|$ would introduce higher order contributions to the effective Hamiltonian, and -- crucially -- eventually lead to a breakdown of the prethermal regime of the Floquet system.
Ultimately, we will vary $\tau$ and $\lambda$ in order to achieve the best performance possible.

To perform the optimization, we transform it into a linear-regression task.
First, we discretize the problem via small equidistant time steps $t_{i=1,...,M}$, with $t_1=0$, $t_M=\tau$. Accordingly, we define the vector $\bs{\Delta_p}=\bigl(\Delta_p(t_1),...,\Delta_p(t_M)\bigr)^T  \in  \mathbb{R}^{M}$ for the discretized detuning profile. 
In addition, we adopt the notation $\bs{N}_0(t) \in \mathbb{C}^{|S|}$ for the relevant subset of matrix elements, written as a column vector of dimension $|S|$. 
Then, we can define an $|S| \times M$ matrix $\underline{N_0} = \bigl(\bs{N}_0(t_1),...,\bs{N}_0(t_M)\bigr) $ that acts as a linear map from a detuning profile $\bs{\Delta_p}$ to an effective Hamiltonian,
\begin{equation} \label{eq:2.3.1}
\bs{\mathcal{H}}[\bs{\Delta}_p] = -\frac{dt}{\tau}\,\underline{N_0} \cdot \bs{\Delta_p}.
\end{equation}
Within this discretization, the solution of the least squares minimization problem defined by \eq{eq:2.3.2} can be given in closed form~\cite{saleh2019theory}. It reads
\begin{equation} \label{eq:2.3.3}
\bs{\tilde{\Delta}}_p = \bs{\tilde{\Delta}}_p(\lambda,\tau,\bs{T}) =  -\biggl( \frac{dt}{\tau}\underline{N_0}^\dagger \underline{N_0} + \tau \lambda \, \mathbb{1} \biggr)^{-1} \underline{N_0}^\dagger \cdot \bs{T},
\end{equation}
and is linear in the target matrix elements $\bs{T}$. 
With the discretized, time-evolved number operators in $\underline{N_0}$ as input, \eq{eq:2.3.3} returns a detuning profile that optimizes the matrix elements of the effective Hamiltonian subject to a finite cost for large detuning perturbations. As the number of discretized steps $M\rightarrow \infty$ ($dt \rightarrow 0$) increases, the optimized profile $\bs{\tilde{\Delta}}_p$ converges to a \textit{continuous} function $\tilde{\Delta}_p(t)$. This property follows from the continuity of the time-evolved operator $\hat{N}_0(t)$, see Appendix~\ref{sec:optimization}.
We note that we may in principle also consider discrete contributions to $\Delta_p(t)$ in \eq{eq:2.2.2}. However, for the remainder of this work, we restrict our use of this possibility to sharp detuning pulses of weight $\Delta_0/\Omega$ at multiples of the period $2\tau$, which gives rise to a detuning contribution $-\frac{\Delta_0}{2\tau\Omega}\hat{N}$ to $\hat{\mathcal{H}}$, see Appendix~\ref{sec:app_Heff}. To first order, the detuning in the effective Hamiltonian $\hat{\mathcal{H}}$ is thus freely adjustable.

In Sec.~\ref{sec:LGTs} we apply this optimization protocol to several examples by evaluating $\hat{N}_0(t) = \sum_i \hat{n}_i(t)$ numerically for small systems. Importantly however, due to the locality of the operators $\hat{n}_i(t\leq \tau)$, the presence of Lieb-Robinson bounds ensures that our approach generalizes to large systems and eventually converges in the thermodynamic limit, see Appendix~\ref{sec:optimization}.

\subsection{Effect of Regularization} \label{sec:regularization}
In order to gain intuition about the optimization procedure, it is helpful to consider the singular value decomposition (SVD) of the operator matrix $\underline{N_0} = W D V^{\dagger}$.
The matrix $W$ is an $|S| \times |S|$ unitary matrix, whose columns form an orthonormal basis for the subspace of \textit{engineerable Hamiltonians} in \eq{eq:2.3.1}.
In contrast, the matrix $V$ is an $M \times M$ real unitary matrix (see Appendix~\ref{app:operator_basis}), whose columns form an orthonormal basis of detuning profiles. Each detuning profile $\Delta_j(t_i) = V_{ij}$ realizes the effective Hamiltonian $(\bs{\mathcal{H}}_j)_l = W_{lj}$ when applied. The proportionality constant is given by the singular values $D_j$, encoded by the diagonal matrix $D$.
As such, the singular value associated with each profile/Hamiltonian pair tells us about the \textit{efficiency} of realizing the corresponding interaction. 
Typically, we find that large singular values arise from smoothly varying profiles, which thus correspond to `easy-to-engineer' interactions. In constrast, small singular values arise from rapidly oscillatory contributions and thus represent interactions that are more challenging to capture. We investigate this effect in more detail in Appendix~\ref{app:operator_SVD}.

Including the parameter $\lambda$ in \eq{eq:2.3.3} then amounts to an effective cutoff on the profiles that contribute to the optimization procedure. When $D_j \ll \lambda$, the contribution of $\bs{\Delta}_j$ to the optimized solution $\bs{\tilde{\Delta}}_p$ is suppressed. This is physically intuitive: Profiles $\bs{\Delta}_j$ with small $D_j$ generate contributions to $\hat{\mathcal{H}}$ with small efficiency, i.e., small prefactor $\sim D_j$.
To make such contributions sizeable, i.e., observable on accessible timescales, one must scale up $\bs{\Delta}_j$, contrary to its role as a perturbation.
This highlights a limitation inherent to our approach: Only target operators $\hat{T}$ contained in the span of profiles with large singular values can be engineered with \textit{both} high fidelity and appreciable prefactor. For target operators outside this span, our regularization can be thought of as biasing the effective Hamiltonian towards the closest efficiently engineerable approximation.

It remains to choose the period $\tau$ and the parameter $\lambda$ that both enter the cost function \eq{eq:2.3.2}. 
When the target matrix elements $\hat{T}$ include multi-body interactions, small evolution times $\tau$ are not sufficient to generate such terms at appreciable weight in $\hat{N}_0(t)$ for $t\leq \tau$. 
On the other hand, for very long times $\tau$ the operator $\hat{N}_0(t)$ is expected to scramble across an exponential number of Pauli strings, thus strongly reducing the weight available in particular terms of interest. 
Similarly, we can formalize our intuition on the role of $\lambda$ presented above: For small $\lambda$, the optimized $\hat{\mathcal{H}}$ is increasingly aligned with the target direction $\hat{T}$. This results in a large value of the quantity
\begin{equation} \label{eq:dd5}
Q^{}_{\measuredangle}(\lambda,\tau) \equiv \frac{(\hat{T}, \hat{\mathcal{H}} [\tilde{\Delta}_p])}{\|\hat{T}\| \, \|\hat{\mathcal{H}}[\tilde{\Delta}_p]\|} \leq 1,
\end{equation}
where $(\cdot,\cdot)$ denotes the operator inner product (with respect to the matrix elements S, see Appendix~\ref{sec:subspaces}).
However, small $\lambda$ also leads to large detuning perturbations, and so the contribution of the desired target $\hat{T}$ to $\hat{\mathcal{H}}$ per amount of perturbation $\|\tilde{\Delta}_p\|$ is small, as quantified by
\begin{equation} \label{eq:dd6}
Q_{\Delta}(\lambda,\tau) \equiv \frac{(\hat{T}, \hat{\mathcal{H}} [\tilde{\Delta}_p])}{\|\hat{T}\| \, \| \tilde{\Delta}_p\|}.
\end{equation}
Hence, there is a tradeoff in the choice of $\lambda$, and we prove in Appendix~\ref{sec:optimization} that indeed $\partial_\lambda Q_{\measuredangle}(\lambda,\tau) \leq 0$ and $\partial_\lambda Q_{\Delta}(\lambda,\tau) \geq 0$ hold.
Therefore, $Q^{}_{\measuredangle}$ and $Q^{}_\Delta$ quantify the \textit{accuracy} and \textit{efficiency} of a given protocol respectively, and give insight into the tradeoff between supressing unwanted terms and achieving fast dynamics on early time scales.
In practice, we may then examine the quantities of \eqs{eq:dd5}{eq:dd6} for a range of values for $\lambda$ and $\tau$ to find suitable choices.

\section{Simulating dynamics of a 2D lattice gauge theory} \label{sec:LGTs}
We now consider the application of our Hamiltonian engineering scheme for the analog simulation of lattice gauge theories. Due to the presence of local gauge constraints, dynamics in lattice gauge theories requires the presence of multi-body interactions beyond two-body terms. In analog schemes, such interactions can be generated in perturbation theory of a small parameter~\cite{Dai_2017ts}. However, if the required perturbative order is high, their tunability relative to lower-order terms in the Hamiltonian will be limited, and their simulation on experimentally accessible time scales might be challenging. 
In the following, we present a concrete example in which our Floquet-approach can overcome these limitations.

\begin{figure}[t]
\centering
\includegraphics[trim={0cm 0cm 0cm 0cm},clip,width=0.99\linewidth]{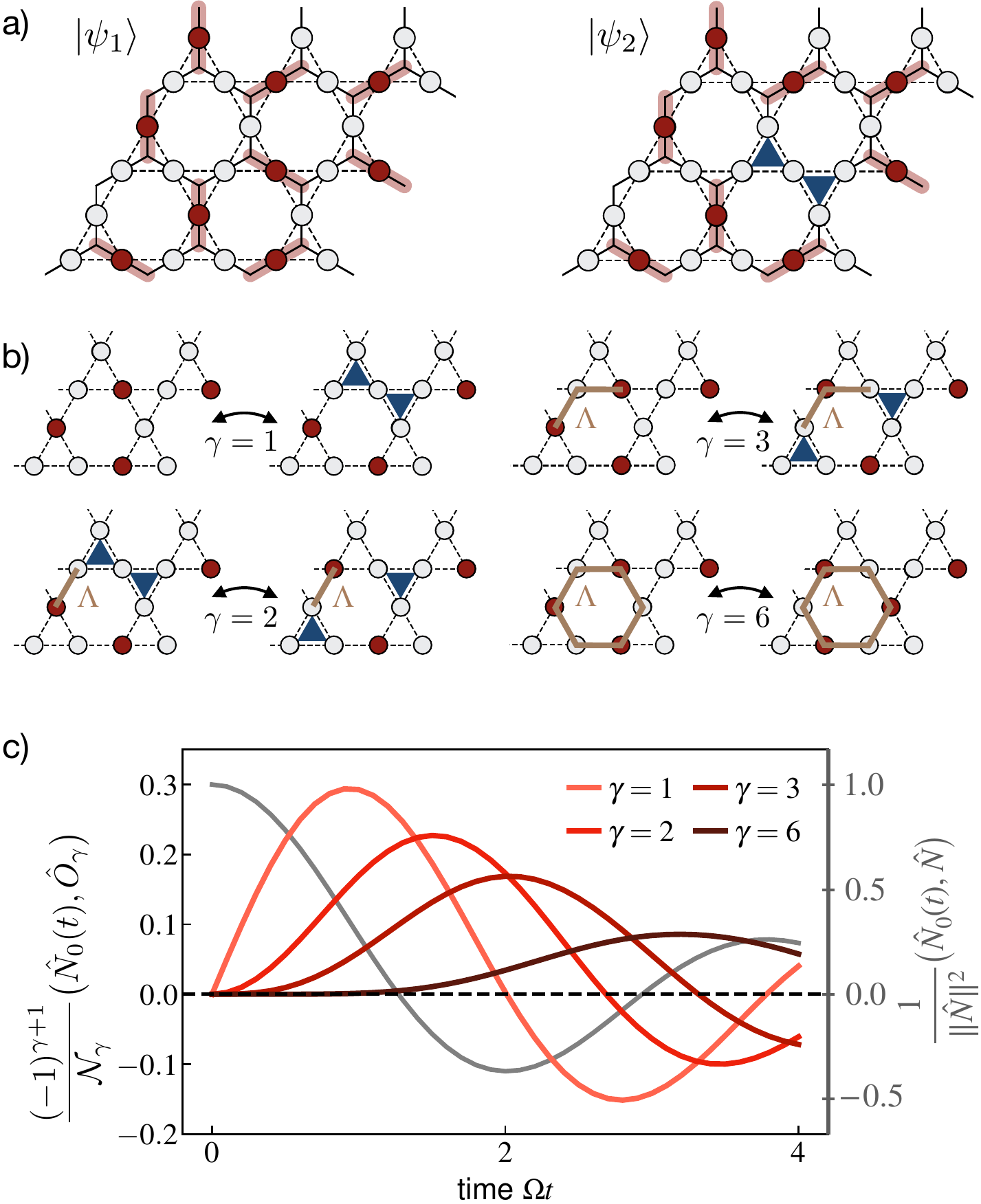}
\caption{\textbf{Kagome lattice PXP model.} \textbf{a)} We numerically study a PXP model on a Kagome lattice of $3\times 3$ unit cells. Each elementary triangle hosts at most one excited spin (full circles) due to nearest neighbor blockade. \textit{Left:} States with maximum number of excited spins map to close-packed dimer states on the honeycomb lattice, yielding a $U(1)$ gauge theory. Shown is a specific example $\ket{\psi_1}$. \textit{Right:} Triangles without excited spins are created in pairs and correspond to charged matter excitations under this $U(1)$ gauge field. The depicted state $\ket{\psi_2}$ contains a pair of charges. \textbf{b)} Dynamical processes of increasing (off-diagonal) operator weight $\gamma$. $\gamma=2$ corresponds to hopping of charges, $\gamma=6$ to plaquette terms that generate pure gauge field dynamics; $\gamma=1$ and $\gamma=3$ create/annihilate pairs of charges. \textbf{c)} Overlap of the time evolved operator $\hat{N}_0(t)$ with the off-diagonal operators illustrated in b). For intermediate times, $\hat{N}_0(t)$ develops significant overlap with off-diagonal processes beyond single spin flips.}
\label{fig:3.1}
\end{figure}

\subsection{Kagome lattice PXP model and $U(1)$ gauge theory}
We consider a PXP model of spins on the sites of the Kagome lattice, see \figc{fig:3.1}{a}. Due to the blockade, each elementary triangle can host at most one site occupying the $n_i=1$ state. The subspace of states with exactly one occupied site on each triangle is exponentially large and equivalent to the close-packed dimer coverings of a honeycomb lattice. The close-packing constraint is equivalent to the condition 
\begin{equation} \label{eq:gl1}
\hat{G}^{\mathrm{(dimer)}}_{\triangle / \bigtriangledown} = 1 - \sum_{i \in \triangle / \bigtriangledown} \hat{n}_i = 0
\end{equation} 
for each triangle of the Kagome geometry. 
Dimer models on bipartite lattices that preserve these close-packing constraints map onto pure $U(1)$ gauge theories~\cite{moessner2001_dimer}. Under this equivalence, the occupation numbers $n_i=0,1$ map onto a spin-$1/2$-valued local electric field $E_i=\pm 1/2$; thus giving rise to quantum link models~\cite{horn1981finite,orland1990lattice,chandrasekharan1997quantum}. Accordingly, the constraints \eq{eq:gl1} correspond to local Gauss laws, with $\hat{G}_{\triangle / \bigtriangledown}$ generating $U(1)$ gauge transformations.

Flipping a single qubit in a close-packed state creates a pair of up- and downward pointing vacant triangles, see \figc{fig:3.1}{a}. In the language of gauge theory, these correspond to positively and negatively charged matter excitations, respectively. Introducing a local occupation number $\hat{\rho}_{\triangle / \bigtriangledown} \in \{0,1\}$ for vacant triangles, the Gauss laws in the presence of charged matter become 
\begin{equation} \label{eq:gl2}
\hat{G}_{\triangle / \bigtriangledown} = 1- \hat{\rho}_{\triangle / \bigtriangledown} - \sum_{i \in \triangle / \bigtriangledown} \hat{n}_i = 0.
\end{equation}  
In the physical setup on the Kagome lattice, the $\hat{\rho}_{\triangle / \bigtriangledown}$ are not independent degrees of freedom, but entirely determined by the electric fields $\hat{n}_i$. Per nearest-neighbor blockade, the Gauss laws \eq{eq:gl2} are thus valid by construction. In this approach of ``integrating out the Gauss laws'', gauge invariance need not be enforced explicitly.
However, to construct a sensible gauge theory, we generally demand that the charged matter excitations have a finite energy gap. In particular, often it is interesting to consider the limit in which the number $\sum_{\triangle} \hat{\rho}_{\triangle} + \sum_{\bigtriangledown} \hat{\rho}_{\bigtriangledown}$ of matter excitations, and thus also $\hat{N}$, is conserved.

One can directly realize this condition in the PXP model at large static detuning, $\Delta(t) = \Delta \gg \Omega$. A Schrieffer-Wolff transformation in the small parameter $\Omega/\Delta$ leads to an effective Hamiltonian~\cite{verresen2021_sl}
\begin{equation} \label{eq:Hstat}
\begin{split}
\hat{H}_{\mathrm{static}} = &-\Delta \hat{N}  -\frac{\Omega^2}{4\Delta} \sum_{|\Lambda|=2} \hat{P} \hat{\sigma}^y_{\Lambda_1}\hat{\sigma}^y_{\Lambda_2} \hat{P} \\ 
&- \frac{3}{32}\frac{\Omega^6}{\Delta^5} \sum_{|\Lambda|=6} \hat{P} \biggl(\prod_{i=1}^6 \hat{\sigma}^y_{\Lambda_i}\biggr) \hat{P} + ...,
\end{split}
\end{equation}
where each term is given to leading order. 
Note that \eq{eq:Hstat} is a perturbative description valid in a \textit{rotated} basis given by the Schrieffer-Wolff transformation. 
In the original basis of $n_i=0,1$ spin states, the conservation of $\hat{N}$ holds only approximately and deteriorates as the ratio $\Omega/\Delta$ increases. 
The second term in \eq{eq:Hstat} sums over all strings of spin flips along connected paths $\Lambda$ of length $|\Lambda|=2$ on the Kagome lattice. This process corresponds to a nearest neighbor hopping of charged matter excitations, see \figc{fig:3.1}{b}. 
The third term in \eq{eq:Hstat} constitutes a six-body plaquette resonance as depicted in \figc{fig:3.1}{b}. This term is the analogue of a magnetic term $\sim B^2$ in the language of gauge theory and generates pure gauge field dynamics in the close-packed dimer model without vacancies.

The lattice gauge theory \eq{eq:Hstat} features dynamical matter and gauge field degrees of freedom but features several significant limitations. Specifically, magnetic plaquette terms occur at sixth order in $\Omega/\Delta$. Thus, the pure gauge dynamics is weak and possibly challenging to observe on moderate time scales. Increasing the ratio $\Omega/\Delta$ enhances such terms, but is detrimental to the approximate conservation of $\hat{N}$ and leads to a high density of virtual charge excitations.  
Moreover, the dynamics of charged matter excitations occurs already at second order in $\Omega/\Delta$ and naturally dominates the six-body plaquette terms. This limits tunability and precludes access to an interesting strong coupling regime where gauge and matter field dynamics occur on comparable timescales. 
One approach to overcome such limitations in analog setups is to encode gauge and matter degrees of freedom separately in the hardware. Given sufficient local control, such a `bottom-up' approach to lattice gauge theories affords more tunability, see e.g.~\cite{banerjee2012_lgt,tagliacozzo2013_opt,zohar2013_qed,
barbiero2019_flgt}, or ~\cite{Homeier_2023vt,halimeh2023spin} for recent proposals. At the same time, gauge invariance needs to be enforced explicitly, for example via large energy penalties~\cite{halimeh2022stabilizing}. We emphasize that there also exists a large body of work using digital approaches~\cite{banuls2020simulating,funcke2023review}. In what follows we present an alternative hardware efficient route towards realization of these ideas.

\subsection{Floquet protocol: Time-evolved operators} 
We now apply the Floquet framework introduced previously to the Kagome lattice PXP model. Concretely, we work with a system of $L \times L = 3 \times 3$ unit cells containing $3L^2$ sites and periodic boundary conditions, shown in \figc{fig:3.1}{c}. The occupation number basis states are denoted as
\begin{equation}
\ket{n} = \ket{n_1,...,n^{}_{3L^2}},
\end{equation}  
and $N_n \equiv \sum_{i=1}^{3L^2}n_i$ is a short hand label for the number of excited spins in $\ket{n}$. Before targeting a specific Hamiltonian engineering goal, we develop a more general intuition on the capabilities of our approach in this setting. For this purpose, we inspect the properties of the time evolved operator $\hat{N}_0(t)$, the central ingredient in our construction. 
Specifically, we are interested in the overlap of $\hat{N}_0(t)$ with the most relevant, off-diagonal processes of the gauge theory framework depicted in \figc{fig:3.1}{b}.
They correspond to hopping of charge excitations and six-body plaquette terms as present also in \eq{eq:Hstat}, as well as single- and three-body spin flip terms that create/annihilate pairs of charge excitations. These processes can be captured via the off-diagonal operators 
\begin{equation} \label{eq:offdiagonal_ops}
\hat{O}_{\gamma} \equiv \sum_{|\Lambda|=\gamma}\, \hat{P} \biggl(\prod_{i=1}^{\gamma}\hat{\sigma}^y_{\Lambda_i}\biggr) \hat{P}.
\end{equation}
In order to evaluate the contribution of these operators to $\hat{N}_0(t)$ numerically, we restrict to the translationally invariant zero momentum sector, spanned by the states 
\begin{equation}
\ket{n_k} = \frac{1}{\mathcal{N}}\sum_{x,y} e^{ik_x x + ik_y y}\, \hat{T}_{x,y} \ket{n},
\end{equation}
for $k=0$. Here, $\hat{T}_{x,y}$ denotes the translation operator by $x,y$ unit cells and $\mathcal{N}$ is a normalization constant. Accordingly, the set of matrix elements between all $k=0$ basis states is given by
\begin{equation}
S = \{ (\ket{n_{k=0}},\ket{n^\prime_{k=0}}) \}.
\end{equation}
Within this set of matrix elements, we subsequently compute the overlap
\begin{equation} \label{eq:op_overlap}
\frac{1}{\mathcal{N}_{\gamma}} \, \bigl(\hat{N}_0(t),\hat{O}_{\gamma} \bigr),
\end{equation}
with the normalization constant
$
\mathcal{N}_{\gamma} = \bigl\| \hat{O}_{\gamma}\bigr\|^2
$.

The result of \eq{eq:op_overlap} for the $3\times 3$ Kagome system is depicted in \figc{fig:3.1}{c}. At very early times $\Omega t \ll 1$, the overlaps grow as $(\hat{N}_0(t),\hat{O}_{\gamma}) \sim t^{\gamma}$ as expected from a small time expansion of $\hat{N}_0(t)$. Beyond perturbative times, the overlaps display oscillatory behavior, with a period of oscillation that increases with the weight $\gamma$ of the corresponding operator. Thus, intuitively, evolution to later times is required for $\hat{N}_0(t)$ to acquire appreciable overlap with larger weight operators. At the same time, the maxima of the overlap decrease for increasing $\gamma$. This agrees with the intuition that $\hat{N}_0(t)$ starts to scramble across many different operators at late times $\Omega t \gg 1$, such that overlaps with particular operators of interest become smaller. As a consequence, overlap with operators of very large weight is necessarily small at all times. We further include the contribution of the single-body operator $\hat{N}$ to $\hat{N}_0(t)$ in \figc{fig:3.1}{c}. 

The key observation of \figc{fig:3.1}{c} is that between very short Floquet periods $\Omega \tau \ll 1$, where multi-body contributions to $\hat{N}_0(t)$ for $t\leq \tau$ are perturbatively suppressed by their operator weight $\gamma$, and long periods $\Omega \tau \gg 1$, where $\hat{N}_0(t\lesssim \tau)$ is scrambled, there exists a non-perturbative regime $\Omega \tau = \mathcal{O}(1)$ of intermediate Floquet periods that allow for flexible Hamiltonian engineering via \eq{eq:2.2.2}. In particular, the structure of $\hat{N}_0(t)$ at intermediate times allows us to find detuning profiles $\Delta_p(t)$ such that contributions of operators with intermediate weights $\gamma \lesssim \mathcal{O}(\Omega \tau)$ to the effective Hamiltonian can be tuned relative to each other. 
We emphasize that the qualitative structure of $\hat{N}_0(t)$ at finite times visible in \figc{fig:3.1}{c} generalizes to other lattice geometries and system sizes.
In the following, we investigate two specific protocols for engineering gauge field dynamics on the Kagome lattice.

\subsection{Floquet protocol for six-body terms} \label{sec:six_body}
In our first example, we directly engineer the required six-body plaquette terms $\hat{O}_{\gamma=6}$, following the scheme developed in Sec~\ref{sec:method}.  
A close-packed dimer model is realized in the sector of maximum Rydberg excitation number $N=L^2$, which our effective Hamiltonian should preserve. 
For this purpose, on top of the continuous detuning $\Delta_p(t)$ derived below, we add a discrete detuning perturbation of strength $\Delta_0/\Omega$ at multiples of $t=2\tau$. This gives rise to a detuning field $-\frac{\Delta_0}{2\tau\Omega}\hat{N}$ in the effective Hamiltonian that stabilizes global Rydberg number, see Appendix~\ref{sec:app_Heff}. 
For concreteness, we fix $\Delta_0/\Omega = 0.7$ in the following.

\begin{figure}[t]
\centering
\includegraphics[trim={0cm 0cm 0cm 0cm},clip,width=0.99\linewidth]{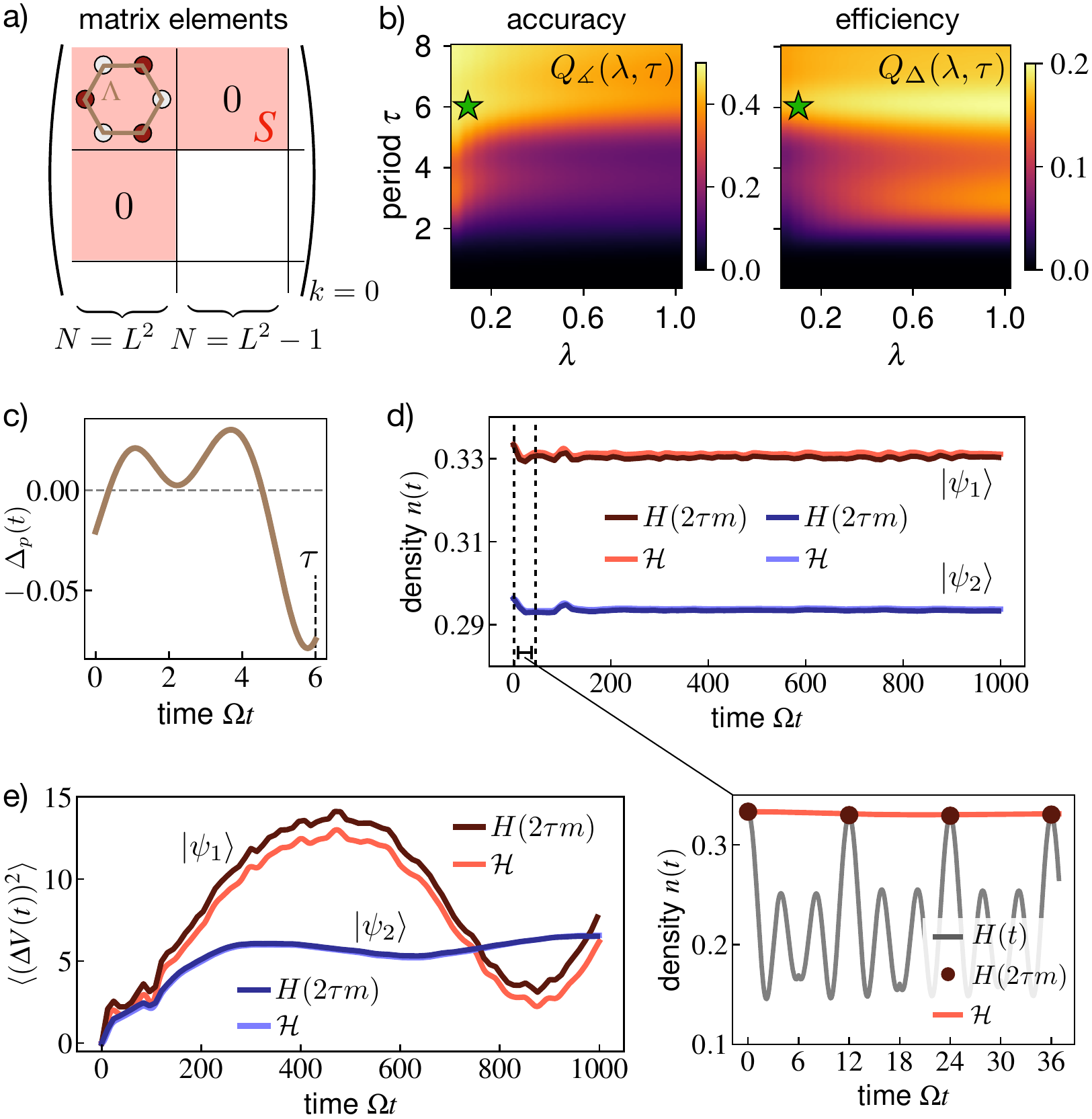}
\caption{\textbf{Engineering six-body terms.} \textbf{a)} We optimize the effective Hamiltonian in the $k=0$ momentum sector with respect to the matrix elements $S$ of \eq{eq:f1_me}. Within the subspace of dimer states with $N=L^2$ excited spins, we optimize towards six-body plaquette terms while minimizing all number-changing matrix elements. \textbf{b)} The optimization is carried out for different values of the Floquet period $\tau$ and the detuning cost $\lambda$. We fix values (marked with a green star) where both the alignment with the target matrix elements $Q_{\measuredangle}(\lambda,\tau)$, as well as the target strength per amount of detuning $Q_{\Delta}(\lambda,\tau)$ for the corresponding effective Hamiltonian are large. \textbf{c)} The resulting, optimized detuning profile for $\Omega \tau = 6$, $\lambda=0.1$. \textbf{d)} Time evolution of the Rydberg number density $n(t)$, starting from $\ket{\psi_1}$ (red) and $\ket{\psi_2}$ (blue), see \figc{fig:3.1}{a}. Under $\Delta_p(t)$ of c) and discrete pulses described in the main text, number conservation is strongly broken in the micromotion (see inset), but highly accurate at multiples of the stroboscopic time $2\Omega \tau = 12$, in agreement with dynamics under the effective Hamiltonian $\hat{\mathcal{H}}$. \textbf{e)} Dynamics of the variance $(\Delta V(t))^2$ of next-nearest-neighbor excitations, \eq{eq:def_variance}. Oscillations in $(\Delta V(t))^2$ for $\ket{\psi_1}$ (red) mark the onset of plaquette dynamics, while $\ket{\psi_2}$ (blue) includes the dynamics of charged matter excitations.
}
\label{fig:f1_opt}
\end{figure}

We then use the optimized continuous part $\Delta_p(t)$ of the detuning profile to construct the six-body plaquette terms. We are mainly interested in the sector of maximum Rydberg occupation number $N=L^2$, and thus restrict our optimization to matrix elements within this sector, as well as matrix elements between this sector and states with occupation number $N=L^2-1$. Moreover, we again work in the translationally invariant sector of zero momentum states. Thus, we define the set of relevant matrix elements to optimize over as
\begin{equation} \label{eq:f1_me}
S = \bigl\{ (\ket{n_{k=0}},\ket{n^\prime_{k=0}}) : N_n + N_{n^\prime} \geq 2L^2 -1 \bigr\}.
\end{equation}
Our target operator within this set of matrix elements is given by
\begin{equation} \label{eq:f1_te}
\hat{T} = g\, \hat{O}_{\gamma=6},
\end{equation}
with prefactor $g$. 
Here, we fix a small $g/\Omega=0.012$ to keep deviations from number conservation small.
Taking \eq{eq:f1_me} and \eq{eq:f1_te} together, our approach targets six-body processes within the $N=L^2$ sector while simultaneously minimizing all matrix elements connecting it to the sector with $N=L^2-1$. An illustration of the matrix elements $S$ and the desired target operator is given in \figc{fig:f1_opt}{a}.

We now determine $\Delta_p(t)$ by minimizing the cost function $C_{\lambda,\tau}(\hat{T})$ of \eq{eq:2.3.2}. To choose the period $\tau$ as well as the detuning cost parameter $\lambda$, we perform this optimization for multiple values of $\lambda,\tau$ and plot the quantities $Q_{\measuredangle}(\lambda,\tau)$ and $Q_{\Delta}(\lambda,\tau)$ defined in \eqs{eq:dd5}{eq:dd6} in \figc{fig:f1_opt}{b}. They quantify the alignment of the resulting effective Hamiltonian with the target direction along $\hat{T}$ and the strength of the effective Hamiltonian along the target direction per amount of detuning perturbation, respectively. Based on the result of \figc{fig:f1_opt}{b}, we select $\Omega \tau = 6.0$ and $\lambda = 0.1$, where both $Q_{\measuredangle}(\lambda,\tau)$ and $Q_{\Delta}(\lambda,\tau)$ are large. The detuning profile $\Delta_p(t)$ resulting from the optimization at these parameter values is shown in \figc{fig:f1_opt}{c}.

With $\Delta_p(t)$ fixed, we consider time evolution under the full, time-dependent Hamiltonian $\hat{H}(t)$ and compare it with the dynamics resulting from our expression for the effective Hamiltonian $\hat{\mathcal{H}}$. Here, we consider $\hat{\mathcal{H}}$ as calculated from \eq{eq:2.2.2} acting on the full $k=0$ Hilbert space. 
\figc{fig:f1_opt}{d} shows the dynamics of the density $n(t) \equiv \braket{\hat{N}}/(3L^2) \leq 1/3$ of excited spins starting from a close-packed dimer initial state. The dynamics under $\hat{H}(t)$ leads the system far away from the initial Rydberg number sector, but comes back to it at multiples of the stroboscopic time $2\tau\Omega = 12$, in agreement with dynamics under the static effective Hamiltonian $\hat{\mathcal{H}}$. The driven system thus approximately conserves the manifold of dimer states at stroboscopic times. 
To verify that non-trivial dynamics occurs within this manifold, we define the number of next-nearest neighbor excited spins,
\begin{equation}
\hat{V} = \sum_{\braket{\braket{i,j}}} \hat{n}_i \hat{n}_j,
\end{equation}
and consider the time evolution of its variance,
\begin{equation} \label{eq:def_variance}
\braket{(\Delta \hat{V}(t))^2} = \braket{\hat{V}^2(t)} - \braket{\hat{V}(t)}^2. 
\end{equation}
Starting from the translationally invariant version of the dimer initial state $\ket{\psi_1}$, $\braket{(\Delta \hat{V}(0))^2} = 0$. As the state turns into a superposition of dimer states with different values of $\hat{V}$, $\braket{(\Delta \hat{V}(t))^2}$ starts to increase on the time scale at which effective plaquette terms occur. \figc{fig:f1_opt}{e} demonstrates oscillating dynamics of $\braket{(\Delta \hat{V}(t))^2}$, with good agreement between the driven time evolution and dynamics under $\hat{\mathcal{H}}$. Simultaneously, the excitation number density remains very high, $n(t) \gtrsim 0.33$, demonstrating that the increasing variance $\braket{(\Delta \hat{V}(t))^2}$ is indeed due to the formation of superpositions in the subspace of dimer states. 
Moreover, we consider dynamics for the initial state $\ket{\psi_2}$ containing a pair of charged matter excitations. Again, we find a high quality of approximate number conservation and excellent agreement between driven evolution and the dynamics of $\hat{\mathcal{H}}$. Non-trivial dynamics of $\braket{(\Delta \hat{V}(t))^2}$ occurs on timescales similar to the dimer initial state.

\begin{figure}[t]
\centering
\includegraphics[trim={0cm 0cm 0cm 0cm},clip,width=0.99\linewidth]{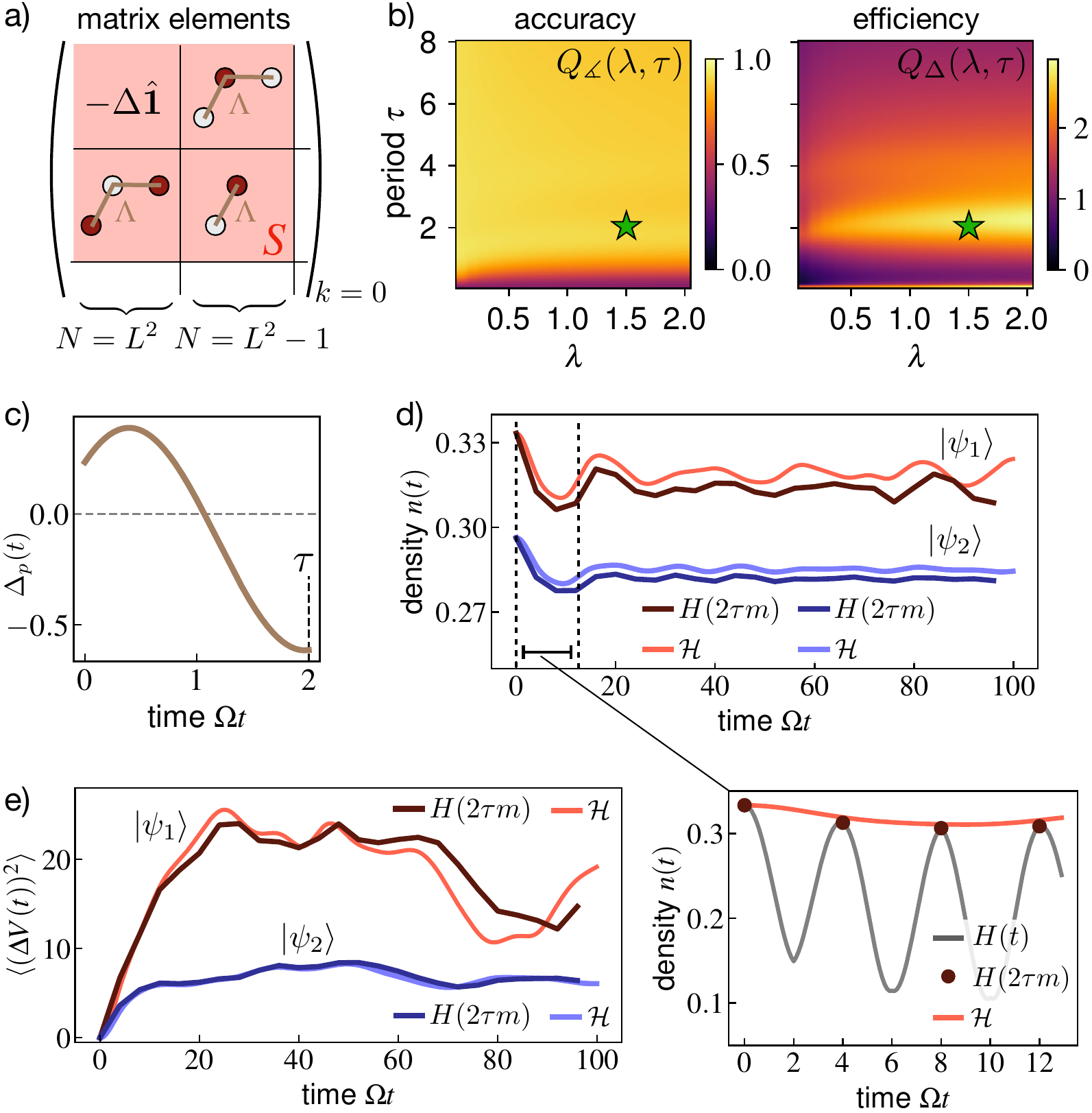}
\caption{\textbf{Engineering three-body terms.} \textbf{a)} Optimization of $\hat{\mathcal{H}}$ in the $k=0$ momentum sector with respect to the matrix elements $S$ of \eq{eq:f2_me}. 
Within the subspace of states with $N=L^2$ or $N=L^2-1$ excited spins, we optimize towards three-body number-changing terms, two-body hopping processes for charged matter excitations, and a detuning contribution.
\textbf{b)} Optimization across different values of period $\tau$ and detuning cost $\lambda$. We fix values marked with a green star, where both $Q_{\measuredangle}(\lambda,\tau)$ and $Q_{\Delta}(\lambda,\tau)$ (see \eqs{eq:dd5}{eq:dd6}) are large. Due to optimization for lower-weight terms compared to \fig{fig:f1_opt}, smaller periods $\tau$ are sufficient.
\textbf{c)} Optimized detuning profile for $\Omega \tau = 2$, $\lambda=1.5$. \textbf{d)} Time evolution of the density $n(t)$ starting from $\ket{\psi_1}$ and $\ket{\psi_2}$ (see \figc{fig:3.1}{a}). Stroboscopic number conservation remains robust, and agreement with effective Hamiltonian dynamics remains accurate at early times. \textit{Inset:} Micromotion of the density in between stroboscopic times.
\textbf{e)} The variance $(\Delta V(t))^2$ shows non-trivial dynamics for both $\ket{\psi_1}$ (red) and $\ket{\psi_2}$ (blue) already on intermediate timescales, indicating the presence of strong effective six-body dynamics induced by three-body interactions.
}
\label{fig:f2_opt}
\end{figure}

Although our approach suppresses terms in $\hat{\mathcal{H}}$ that do not commute with $\hat{N}$, they generally remain non-zero.
To verify that dynamics of $\hat{\mathcal{H}}$ in the fully-packed manifold is indeed dominated by six-body interactions, we consider a modified version of the effective Hamiltonian (in the $k=0$ sector):
\begin{equation} \label{eq:f1_scale}
\begin{split}
\hat{\mathcal{H}}(\nu) \equiv \nu \hat{\mathcal{H}} + \frac{1-\nu}{\|\hat{N}\|^2} \bigl(\hat{N},\hat{\mathcal{H}}\bigr) \, \hat{N}.
\end{split}
\end{equation}
$\hat{\mathcal{H}}(\nu)$ rescales the part of $\hat{\mathcal{H}}$ orthogonal to $\hat{N}$ by a factor $\nu$. 
To analyze the effect of this rescaling on the dynamics of $\braket{(\Delta \hat{V}(t))^2}$, starting from $\ket{\psi_1}$, we consider the frequency $\omega_{\mathrm{max}} \equiv \max_{\omega \neq 0} \bigl|\mathcal{F}[\braket{(\Delta \hat{V}(t))^2}](\omega)\bigr|$ at which the Fourier transform $\mathcal{F}[\braket{(\Delta \hat{V}(t))^2}](\omega)$ of the time-evolved variance is maximal. For dominant six-body terms, we expect a \textit{linear} relation $\omega_{\mathrm{max}}\sim \nu$ upon rescaling, and our numerics are consistent with this expectation as shown in \figc{fig:scaling}{b}. This further becomes manifest in a linear scaling collapse of the early time dynamics of $\braket{(\Delta \hat{V}(t))^2}$, \figc{fig:scaling}{c}. At the same time, the long-time average $\overline{n(t)}$ of the density remains near its maximum value, see \figc{fig:scaling}{a}.  
By contrast, applying an analogous rescaling to a PXP model at static detuning results in a sixth-order relation $\omega_{\mathrm{max}} \sim \nu^6$ and leads to a rapidly changing average density $\overline{n(t)}$, see \figc{fig:scaling}{a,b}. 

Overall, the protocol of this section induces dynamics dominated by six-body plaquette terms at a very high quality of number conservation. In the following, we attempt to further enhance the strength of effective six-body dynamics to enable evolution on even faster time scales.

\subsection{Floquet protocol for three-body terms} \label{sec:three_body}
In this section, we follow a different route towards effective plaquette terms: Instead of generating six-body interactions directly, we engineer an effective Hamiltonian in which the pure gauge field dynamics is predominantly generated via second-order perturbative processes from three-body terms. 
As in the previous example, we employ a discrete pulse at multiples of the period $2\tau$ to generate a discrete contribution
$
-\frac{\Delta_0}{2\tau\Omega} \hat{N}
$ to the effective Hamiltonian, fixing $\Delta_0/\Omega = 1.0$ in the following.

For the optimization of the continuous part of the detuning profile, we take into account all matrix elements between $k=0$ states with either $N=L^2$ or $N=L^2-1$, i.e.,
\begin{equation} \label{eq:f2_me}
S = \bigl\{ (\ket{n_{k=0}},\ket{n^\prime_{k=0}}) : N_n, N_{n^\prime} \geq L^2-1 \bigr\}.
\end{equation}
We now define our target operator within this set of matrix elements as
\begin{equation} \label{eq:f2_te}
\hat{T} = -\Delta \hat{N} + g^{}_3 \, \hat{O}_{\gamma=3} + g^{}_2 \, \hat{O}_{\gamma=2},
\end{equation}
where we set $\Delta / \Omega = 0.22$, $g_3/\Omega = -g_2/\Omega = 0.055$. Hence, $\hat{T}$ gives an additional contribution to the detuning and balances the two- and three-body terms corresponding to the processes shown in \figc{fig:3.1}{b}. Crucially, $\hat{T}$ does not contain single-body spin flip terms. We illustrate the matrix elements and target operator entering our optimization scheme in \figc{fig:f2_opt}{a}. 
Next, we perform the optimization and evaluate the quantities $Q_{\measuredangle}(\lambda,\tau)$, $Q_{\Delta}(\lambda,\tau)$ for multiple values of $\lambda$, $\tau$ in \figc{fig:f2_opt}{b}. Both are found to be large for $\Omega \tau = 2.0$, $\lambda = 1.5$. The optimized detuning profile corresponding to these parameter values is shown in \figc{fig:f2_opt}{c}. The chosen Floquet period is significantly shorter compared to the previous section, a consequence of optimizing for operators with smaller weight. 
Moreover, \figc{fig:f2_opt}{b} shows that $\hat{\mathcal{H}}$ is well-aligned with the target direction along $\hat{T}$, $Q_{\measuredangle}(\lambda=1.5,\Omega\tau=2.0) \approx 0.9$.

With $\Delta_p(t)$ fixed, we show the dynamics from the state $\ket{\psi_1}$ under $\hat{H}(t)$ in \figc{fig:f2_opt}{d}. At stroboscopic times, the approximate conservation law is less pronounced compared to the previous section, but $n(t) \gtrsim 0.31$ remains very high, consistent with evolution under $\hat{\mathcal{H}}$.
This observation extends to the variance $(\Delta V(t))^2$ of \eq{eq:def_variance}, which exhibits oscillations, and thus non-trivial gauge field dynamics, already on the times $\Omega t \leq 20$ shown in \figc{fig:f2_opt}{e}. Dynamics from the state $\ket{\psi_2}$ with a pair of charged matter excitations occurs on similar time scales.

\begin{figure}[t]
\centering
\includegraphics[trim={0cm 0cm 0cm 0cm},clip,width=0.99\linewidth]{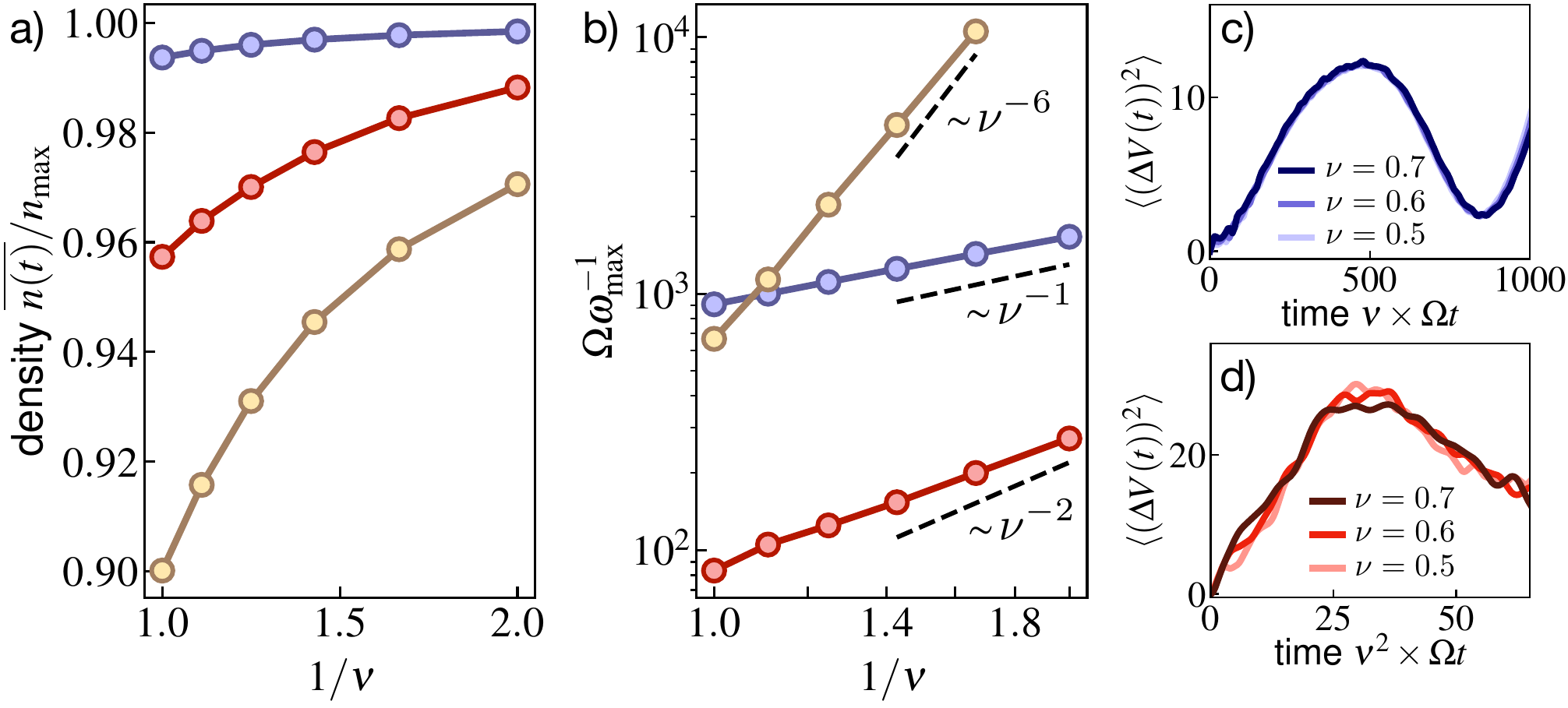}
\caption{\textbf{Effective Hamiltonian dynamics.} Dynamics of dimer state $\ket{\psi_1}$ under $\hat{\mathcal{H}}(\nu)$ of \eq{eq:f1_scale}, which rescales the all terms in $\hat{\mathcal{H}}$ orthogonal to $\hat{N}$ by a factor $\nu$. \textbf{a)} Long-time average $\overline{n(t)}$ of the density as a function of $\nu$ for the six-body protocol outlined in Sec.~\ref{sec:six_body} (blue), the three-body protocol of Sec.~\ref{sec:three_body} (red), and a static PXP model at $\Delta/\Omega = 2.0$ (yellow). \textbf{b)} Dominant frequency $\omega_{\mathrm{max}}$ in the dynamics of the variance $(\Delta V(t))^2$. The observed scalings $\omega_{\mathrm{max}} \sim \nu,\, \nu^2, \, \nu^6$ are consistent with dominant six-body, three-body and single-body terms, respectively. \textbf{c)+d)} Scaling collapse of early time dynamics of $(\Delta V(t))^2$ for six- and three-body protocols. 
}
\label{fig:scaling}
\end{figure}

As in the previous section, we want to verify that the gauge field dynamics of $\hat{\mathcal{H}}$ is indeed dominated by low-order perturbative processes.
For this purpose, we consider the rescaling of all terms orthogonal to $\hat{N}$ as in \eq{eq:f1_scale}. Accordingly, for strong second-order processes, we see the dominant oscillation frequency $\omega_{\mathrm{max}}$ of the variance $(\Delta V(t))^2$ scale approximately as $\omega_{\mathrm{max}}\sim \nu^2$, \figc{fig:scaling}{b}. This relation becomes manifest in the early time dynamics of $(\Delta V(t))^2$ upon rescaling the time axis with the factor $\nu^2$. We further see in \figc{fig:scaling}{a} that the long time average $\overline{n(t)}$ varies more strongly with $\nu$ compared to the six-body case of the previous section, but remains well above the static PXP case despite orders-of-magnitude more efficient gauge field dynamics.

Finally, we point out the general tradeoff between accuracy and efficiency of implementing the targeted dynamics in \fig{fig:scaling}: As the strength $\nu$ in \eq{eq:f1_scale} becomes stronger, terms that were not optimized for are scaled up as well. Thus, while dynamics becomes faster, \figc{fig:scaling}{b}, its accuracy decreases as shown for example in the decreasing quality of number conservation in the six-body protocol, \figc{fig:scaling}{a}, which was initially optimized for ideal number conservation.
In addition, there may be processes besides those specifically targeted that contribute to the gauge field dynamics, such as third order processes. We leave as an interesting open question whether an even better tradeoff between accuracy and efficiency for realizing pure gauge field dynamics can be constructed by specifically targeting such processes as well.

\begin{figure*}[t]
\centering
\includegraphics[trim={0cm 0cm 0cm 0cm},clip,width=0.99\linewidth]{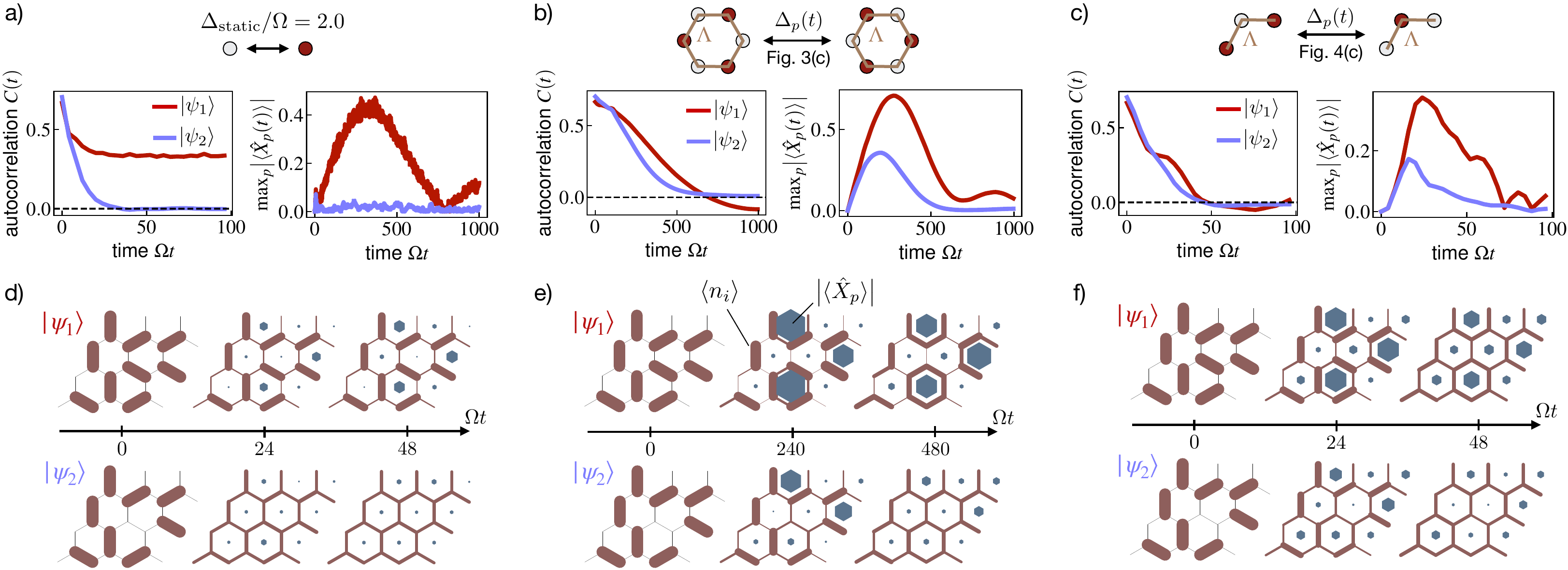}
\caption{\textbf{Local quantum dynamics.} \textbf{a)-c)} (left panels) We characterize the time evolution of local observables for a constant static detuning $\Delta_{\mathrm{static}}/\Omega = 2.0$ and the Floquet protocols of Sec.~\ref{sec:six_body} and Sec.~\ref{sec:three_body}. In the static setting a), the autocorrelation $C(t)$ (see \eq{eq:autocorr}) of local densities exhibits a large separation of timescales between the fully packed dimer initial state $\ket{\psi_1}$ and the initial state $\ket{\psi_2}$ containing a pair of charge excitations. In contrast, the autocorrelation function of both initial states decays on similar timescales in our Floquet protocols b)+c), indicating comparable coupling strengths of pure gauge and matter dynamics. 
\textbf{a)-c)} (right panels) Time evolution of the strongest plaquette resonance $\max_p |\braket{\hat{X}}_p|$ in the system (see \eq{eq:plaquette_op}). For the dimer initial state $\ket{\psi_1}$, signifcant resonances eventually develop in all protocols. For $\ket{\psi_2}$, the resonances are lost due to dominant matter hopping processes in the static case a), but persist in the strongly coupled Floquet schemes of b) and c). 
\textbf{d)-f)} Site-resolved dynamics of densities and plaquette expectation values in the protocols of a)-c), respectively. The thickness of the bonds indicates the local Rydberg density $\braket{\hat{n}_i}$, the area of the blue hexagons the local plaquette resonance expectation value $|\braket{\hat{X}_p}|$.
In the static protocol d), the local densities show a strong memory of the dimer initial state $\ket{\psi_1}$, while the state $\ket{\psi_2}$ rapidly becomes homogeneous due to dominant matter dynamics. In contrast, in the Floquet approaches e) and f), both states become homogeneous on roughly similar time scales and develop plaquette resonances even in the presence of matter excitations. The protocol shown in f) achieves this on relatively short time scales.
}
\label{fig:local_dyn}
\end{figure*}

\subsection{Interplay of gauge and matter dynamics} \label{sec:interplay}
We next investigate the physical consequences of the strong plaquette interactions generated via our Floquet approach. In particular, we are interested in the interplay between dynamical gauge and matter degrees of freedom in this strong coupling regime.
For this purpose, we first examine the local relaxation dynamics starting from the non-translationally invariant versions (in contrast to the previous sections) of occupation number initial states $\ket{\psi}$. To quantify their relaxation, we consider the autocorrelation function 
\begin{equation} \label{eq:autocorr}
C_\psi(t) \equiv \frac{1}{N_\psi}\sum_i \bigl[\braket{\psi|\hat{n}_i(t)|\psi}- \braket{\hat{n}_i}_{N_\psi}\bigr] \braket{\hat{n}_i}_\psi.
\end{equation}
Here, $\braket{\hat{n}_i}_\psi = \braket{\psi|\hat{n}_i|\psi}$, and $\braket{\hat{n}_{i}}_{N_\psi}$ denotes the average of $\hat{n}_i$ over the ensemble of all product states $\ket{n}$ with total occupation number $N_n = N_\psi$ equal to the initial state. In a thermalizing system at high temperature but with number conservation, we expect $C_\psi(t\rightarrow \infty) = 0$ at long times. We show $C_\psi (t)$ starting from $\ket{\psi_1}$ and $\ket{\psi_2}$ containing no charge excitations and a pair of charges, respectively, in \figc{fig:local_dyn}{a-c}. Crucially, we observe that while a static approach leads to a large separation of timescales in the decay of $C_{\psi_1}(t)$ and $C_{\psi_2}(t)$ (\figc{fig:local_dyn}{a}), both initial states relax on comparable times in the Floquet protocols introduced in Sec.~\ref{sec:six_body} and Sec.~\ref{sec:three_body} (\figc{fig:local_dyn}{b-c}). The more oscillatory character of $C_{\psi_1}(t)$ compared to $C_{\psi_2}(t)$ is likely due to the smaller Hilbert space of fully packed dimer states, and thus a finite size effect.
In Appendix~\ref{sec:snapshots}, we further show that computational basis snapshot measurements provide a qualitative view into the mechanism by which gauge field dynamics is generated in each protocol.

For initial states containing matter excitations, the balance between gauge and matter field couplings results in a dynamical competition that can be diagnosed by the local plaquette operators
\begin{equation} \label{eq:plaquette_op}
\hat{X}_p \equiv \hat{\sigma}^x_{p_1}\hat{\sigma}^y_{p_2} \hat{\sigma}^y_{p_3} \hat{\sigma}^y_{p_4} \hat{\sigma}^y_{p_5} \hat{\sigma}^y_{p_6}.
\end{equation}
We note that $\hat{X}_p$ is different from the operators $\hat{O}_{\gamma=6}$ appearing in the effective Hamiltonian. In particular, the presence of $\hat{O}_\gamma$ in $\hat{\mathcal{H}}$ will generate a large expectation value of $\hat{X}_p$ for plaquettes initially in a flippable product state configuration.
By evaluating the strongest plaquette resonance in the system, $\max_{p}|\braket{\hat{X}_p}(t)|$, we show in \figc{fig:local_dyn}{a-c} (right panels) that this is indeed the case in the dynamics from the fully packed dimer initial state $\ket{\psi_1}$. However, starting from the state $\ket{\psi_2}$ with matter excitations, no such plaquette resonances build up for quenches under a static detuning, \figc{fig:local_dyn}{a}.
This is due to charge hopping dynamics dominating the weak pure gauge interactions. 
In contrast, starting from $\ket{\psi_2}$ in our Floquet protocols, \figc{fig:local_dyn}{b,c} demonstrate that significant plaquette resonances still appear at early times before eventual relaxation to a homogenous state. 
This picture is further substantiated by the real-space resolved dynamics of the local densities $\hat{n}_i$ and plaquette resonances $\hat{X}_p$ displayed in \figc{fig:local_dyn}{d-f}.
In the static setting, \figc{fig:local_dyn}{d}, $\ket{\psi_1(t)}$ exhibits a strong memory of the initial state $\ket{\psi_1}$ at times where $\ket{\psi_2(t)}$ quickly relaxes to a homogenous distribution of $\braket{\hat{n}_i}$. $\ket{\psi_2(t)}$ does not feature significant plaquette resonances $\braket{\hat{X}_p}$ at any location in the system. 
In the Floquet approach, \figc{fig:local_dyn}{e,f}, $\ket{\psi_1(t)}$ develops strong resonances for plaquettes initially in a flippable configuration and reaches a homogenous distribution of $\braket{\hat{n}_i}$ at similar times as $\ket{\psi_2(t)}$, see in particular \figc{fig:local_dyn}{f}. $\ket{\psi_2(t)}$ still develops significant resonances $\braket{\hat{X}_p}$ at plaquettes initally in a flippable configuration, demonstrating the competition between gauge and matter dynamics.

\section{Prospects for experimental realizations}
The scheme developed in our work relies on relatively simple control provided by a global detuning field, which is readily available in state-of-the-art Rydberg quantum simulators. In addition, we have identified promising protocols, such as the three-body scheme investigated in Sec.~\ref{sec:three_body}, which display non-trivial gauge field dynamics already at times of a few decades in units of $1/\Omega$. Assuming a Rabi frequency of $\Omega = 2\pi \times 3.0 \mathrm{MHz}$ and a coherence time of $\tau_c = 1\mu\mathrm{s}$, these interactions, as well as their interplay with dynamical matter degrees of freedom, are accessible with present-day hardware. Moreover, as pointed out in Secs.~\ref{sec:regularization},\ref{sec:three_body}, our approach exhibits a tradeoff between supressing unwanted terms and generating fast dynamics, such that the targeted processes may be enhanced to even earlier times by partially relaxing the high quality of Rydberg number conservation.
A more detailed analysis of the effects of noise and other experimental imperfections on our Floquet protocol, in particular for potential heating, is left for future work.

In addition, while we consider a nearest-neighbor blockaded PXP-setup throughout most of this work, a significant aspect of realistic hardware is the presence of long-range tails in the van-der-Waals interactions of the Rydberg Hamiltonian $\hat{H}_{\mathrm{Ryd}}$,
\begin{equation} \label{eq:vdW_int} 
\begin{split}
\hat{H}_{\mathrm{Ryd}}(t) &= \frac{\Omega}{2}\sum_i \hat{\sigma}^x_i - \Delta(t) \hat{N} + \hat{V}_{\mathrm{vdW}} \\
\hat{V}_{\mathrm{vdW}} &= \sum_{i,j} V_{ij} \, \hat{n}_i \hat{n}_j = \sum_{i,j}\Omega\biggl(\frac{R_b}{r^{}_{ij}}\biggr)^6 \hat{n}_i \hat{n}_j,
\end{split}
\end{equation} 
where $R_b$ denotes the blockade radius. When contributions to $V_{\mathrm{vdW}}$ from sites $i,j$ with $r_{ij}>R_b$ are sufficiently small, such long-range terms can be incoporated as additional interactions (conjugated by PXP-evolution) in our formulation of the effective Hamiltonian. They can furthermore be partially counteracted with a constant mean field shift of the global detuning~\cite{bluvstein2021_scars}. 
However, while contributions from long-range tails may be reduced by lowering the blockade radius $R_b$, this also leads to a decrease of the energy penalty for nearest-neighbor blockade violations, which constitute another source of perturbation to our scheme.
Thus, upon tuning $R_b$, there is a tradeoff in the severity of long-range tails and blockade violations, and both must be considered jointly in a realistic treatment of our approach for experimental setups. 
We further emphasize that these perturbations depend on the geometry of the setup and we generally expect them to be smaller in one dimension and on two-dimensional lattices with a small ratio of nearest- to next-nearest-neighbor distances~\cite{bluvstein2021_scars}. This suggests that the hexagonal lattice and the Kagome geometry considered in this work are good candidates for two-dimensional realizations.

Finally, although we previously considered infinitely sharp detuning pulses, we should use finite-width pulses in experimental settings. Small deviations from the idealized delta-shape can again be straightforwardly incorporated as contributions to the effective Hamiltonian in our approach, and will depend both on the location and precise shape of the pulses. A realistic pulse duration for experimental realizations ranges around $\tau_{\mathrm{pulse}}=20\mathrm{ns}$. With the above Rabi frequency, $\Omega \tau_{\mathrm{pulse}} \approx 0.4$, and the pulse width does account for a sizeable fraction of the Floquet period $2\Omega\tau = 4.0$ chosen in our three-body protocol. This suggests that the broadened pulses will contribute substantially to the resulting effective Hamiltonian. However, strong contributions from broadened pulses do not \textit{per se} run counter to our approach: For pulses at even multiples $2m\tau$ of the pulse period $\tau$, deviations from a delta-shape primarily induce contributions to the effective Hamiltonian from $\hat{N}_0(t)$ within a pulse-width of time zero. Per \figc{fig:3.1}{c}, $\hat{N}_0(0 < t < \tau_{\mathrm{pulse}}/2)$ is dominated by the number operator $\hat{N}$, leading to an effective detuning contribution that is desirable in our approach, as it enforces Rydberg number conservation. Moreover, for the three-body protocol with $\Omega\tau = 2.0$, broadened pulses at odd multiples $(2m+1)\tau$ of $\tau$ yield contributions of the operators 
$\hat{N}_0(\tau - \tau_{\mathrm{pulse}} /2 < t < \tau)$
%$\hat{N}_0(\tau - t \lesssim \tau_{\mathrm{pulse}})$
to $\hat{\mathcal{H}}$
. \figc{fig:3.1}{c} shows that such terms are associated with sizeable three- and two-body interactions but only small single spin-flip contributions; they can thus be used to realize the strong three-body dynamics targeted in Sec.~\ref{sec:three_body}. 
Therefore, we may partially view these experimental demands as additional restrictions on the echo-perturbing detuning profiles $\Delta_p(t)$ that can be implemented in practice. Integrating these experimental restrictions into our optimization procedure is an important avenue for future work.
In addition, we note that a finite pulse duration may further be helpful to prevent an abundance of blockade violations, as sharp pulses contain high-frequency components that may resonantly couple to blockade-violating states.
A more detailed analysis that combines our approach with $\hat{H}_{\mathrm{Ryd}}$ of \eq{eq:vdW_int} beyond the approximation of a blockaded Hilbert space and with the full set of experimental constraints on the time-dependent detuning $\Delta(t)$ is left for future work.

\section{Discussion \& Outlook}
We have introduced and analyzed a new method for realizing dynamical gauge theories with strong multi-body interactions in periodically driven systems with nearest-neighbor Rydberg blockade. The key ingredient is a many-body echo realized upon applying a simple global detuning pulse. The micromotion between pulses is then used a resource for operator spreading, which contributes tunable multi-body interactions to an effective Hamiltonian. The latter can be steered towards a given target upon selecting suitable perturbations around the ideal echo point. Using this approach, we constructed effective Hamiltonians of a two-dimensional $U(1)$ lattice gauge theory and demonstrated access to previously unexplored regimes. In particular, this allows us to combine a high quality of particle number conservation with strong six-body magnetic plaquette terms on par with two-body particle hopping processes, a constellation that is otherwise challenging to realize in analog simulation. 

Our approach builds on a number of key concepts in the field of quantum many-body dynamics, such as Floquet engineering, operator spreading, or prethermalization, and exhibits connections to quantum many-body scars and discrete time crystals~\cite{maskara2021_dtc,khemani2016_dtc,else2016_dtc,yao2017_dtc}. On the one hand, it yields new insights on the dynamics of periodically driven many-body systems far from equilibrium. On the other, it combines these concepts to create a new quantum simulation tool that extends the capabilities of systems with Rydberg blockade interactions and enables experimental applications.

Our results can be extended along several promising directions. In particular, the many-body echo considered here is only one of (infinitely) many possible approaches.  We recall that \eq{eq:2.2.2} considers only profiles $\Delta_p(t)$ symmetric around $t=\tau$. Contributions to $\Delta_p(t)$ that are \textit{antisymmetric} around $\tau$ may be absorbed into a new echo evolution $\hat{U}_e^\prime(t)$, leading to differently time-evolved operators in $\hat{\mathcal{H}}$ upon introducing perturbations. 
In addition, the timing of $\pi$-pulses as in \figc{fig:1}{b} may be shifted, giving rise to effective models $\hat{\mathcal{H}} \sim \int_{-\tau_-}^{\tau_+} dt\, \Delta_p(t) \hat{N}_0(t)$ with $\hat{N}_0(t)$ evaluated at negative times. Moreover, larger system sizes and longer evolution times may be analyzed by combining our approach with tensor network methods~\cite{verstraete2008_mps,schollwoeck2011_dmrg,orus2014_mps,hauschild2018_tenpy} to compute the time-evolved local operators $\hat{n}_i(t)$, for example as matrix product operators. 
Furthermore, we note that our scheme applies equally to systems with Rydberg blockade beyond nearest neighbors and systems with site-dependent detuning fields~\cite{manovitz2024quantum}, which allows for the simulation of lattice gauge theories beyond the example studied in this work.

While we focused on nonequilibrium dynamics of the resulting effective Hamiltonian for high-energy initial states, future work may address ground state properties or exotic low-energy dynamics of models that can arise in our Floquet protocol~\cite{lan2018_glassy,feldmeier2019_glassy}. In particular, it would be interesting to determine the presence of ground state phases that are stabilized by multi-body ring-exchange terms, such as plaquette valence bond solids~\cite{moessner2001_hexagon} or even topological spin liquid phases~\cite{misguich2002_sl,verresen2021_sl,semeghini2021_sl,Kornjaca2023_sl}. A related question concerns the preparation of low energy states of the effective Hamiltonian, potentially by exploring connections to counterdiabatic driving schemes~\cite{berry2009transitionless,sels2017_cd,claeys2019_counter}. 
In addition, our protocol unlocks the simulation of blockaded systems with conserved Rydberg number more generally. In Ref.~\cite{koyluoglu2024_ent} we show that this allows for probing gapless phases of matter and the construction of Hamiltonians that generate multi-partite entanglement starting from simple product initial states.

Finally, we note that  our method involving the perturbation around periodically reviving many-body trajectories can be applied to other experimental systems beyond Rydberg atom arrays. Time-periodic many-body dynamics is commonly employed in dipolar-interacting quantum systems~\cite{wahuha1968,wei2018_nmr,choi2020_pulses,zhou2020_metro,geier2024time}, and can be implemented, for instance, in neutral atoms in optical lattices and cavities~\cite{linnemann2016_tr,Colombo2022_tr}, in trapped ions~\cite{gaerttner2017_tr,gilmore2021_tr}, or digitally in superconducting devices~\cite{blok2021_scramble,mi2022_tr,Braumuller2022_tr}. Furthermore, state-dependent periodic revivals occur in systems with quantum-many body scars~\cite{Bernien2017_51atom,bluvstein2021_scars}. Devising Hamiltonian engineering protocols for such systems, analogous to the present work, is a promising direction for future work.

\acknowledgements
We thank Dolev Bluvstein, Sepehr Ebadi, Simon Evered, Alexandra Geim, Sophie Li, Marcin Kalinowski, Tom Manovitz, Roger Melko, Simone Notarnicola, Hannes Pichler, Maksym Serbyn, Norman Yao, and Peter Zoller for insightful discussions.
We acknowledge financial support from the US Department of Energy (DOE Gauge-Gravity, grant number DE-SC0021013, and DOE Quantum Systems Accelerator, grant number DE-AC02-05CH11231), the National Science Foundation (grant number PHY-2012023), the Center for Ultracold Atoms (an NSF Physics Frontiers Center), the DARPA ONISQ program (grant number W911NF2010021), the DARPA IMPAQT program (grant number HR0011-23-3-0030), and the Army Research Office MURI (grant number W911NF2010082).
% PERSONAL FUNDING 
J.F. acknowledges support from the Harvard Quantum Initiative Postdoctoral Fellowship in Science and Engineering.
N.M. acknowledges support by the Department of Energy Computational Science Graduate Fellowship under award number DE-SC0021110.
N.U.K. acknowledges support from The AWS Generation Q Fund at the Harvard Quantum Initiative.

\begin{appendix}

\section{Derivation and applicability of effective Hamiltonian} \label{sec:app_Heff}
We present the derivation of the effective Hamiltonian of \eq{eq:2.2.2}, largely following the works of Refs.~\cite{else2017_dtc,maskara2021_dtc}. Our starting point is the many-body echo described in Sec.~\ref{sec:echo}, consisting of evolution under $\hat{H}_0$ together with periodic detuning pulses $\Delta(t) = \pi \sum_{m \in \mathbb{N}} \delta(t-m\tau)$. We denote the Hamiltonian generating this echo evolution as
\begin{equation} \label{eq:app_echo}
\hat{H}_e(t) = \hat{H}_0 - \pi \hat{N} \sum_{m \in \mathbb{N}} \delta(t-m\tau),
\end{equation}
such that the full time-dependent Hamiltonian is given by 
\begin{equation}
\hat{H}(t) = \hat{H}_e(t) - \Delta_p(t)\hat{N},
\end{equation}
with the detuning perturbation $\Delta_p(t)$. The unitary generated by $\hat{H}_e(t)$ is denoted as $\hat{U}_e(t)=\hat{\mathcal{T}}\exp\bigl(-i\int_0^t dt^\prime \hat{H}_e(t^\prime)\bigr)$ and is explicitly given by
\begin{equation} \label{eq:app1}
\begin{split}
&\hat{U}_e(0 \leq t < \tau) = e^{-it \hat{H}_0} \\
&\hat{U}_e(\tau \leq t < 2\tau) = e^{-i(t-\tau) \hat{H}_0}e^{-i\pi \hat{N}} e^{-i\tau \hat{H}_0} \\
&\hat{U}_e(2\tau) = e^{-i\pi \hat{N}} e^{-i\tau \hat{H}_0}e^{-i\pi \hat{N}} e^{-i\tau \hat{H}_0} = \hat{\mathbb{1}}.
\end{split}
\end{equation}
By definition, the $\pi$-pulses are applied infinitesimally prior to the times $m\tau$, with $m \in \mathbb{N}$.

We now switch to an interaction picture with respect to $\hat{U}_e(t)$, in which states evolve as $\ket{\psi_{\mathrm{I}}(t)} = \hat{U}_e^\dagger(t) \ket{\psi(t)}$ and operators as $\hat{A}_{\mathrm{I}}(t) = \hat{U}_e^\dagger(t) \hat{A}(t) \hat{U}_e(t)$. In this interaction picture, the time evolution of $\ket{\psi_{\mathrm{I}}(t)}$ is readily shown to be generated by the Schrödinger equation
\begin{equation}
\partial_t \ket{\psi_{\mathrm{I}}(t)} = i \Delta_p(t) \hat{N}_{\mathrm{I}}(t) \ket{\psi_{\mathrm{I}}(t)},
\end{equation}
which is formally solved by $\ket{\psi_{\mathrm{I}}(t)} = \hat{\mathcal{T}} \exp \bigl( i \int_0^t dt^\prime \Delta_p(t^\prime) \hat{N}_{\mathrm{I}}(t^\prime) \bigr) \ket{\psi_{\mathrm{I}}(0)}$. We now make use of the echo property $\hat{U}_e(2m\tau) = \hat{\mathbb{1}}$ for $m \in \mathbb{N}$, which implies that $\ket{\psi_{\mathrm{I}}(2m\tau)}=\ket{\psi(2m\tau)}$, i.e., interaction picture and Schrödinger picture coincide at multiples of $2\tau$. Furthermore, $\hat{N}_{\mathrm{I}}(t+2\tau) = \hat{N}_{\mathrm{I}}(t)$ is $2\tau$-periodic.
As a consequence, the evolution at stroboscopic times in the Schrödinger picture is given by
\begin{equation} \label{eq:app2}
\begin{split}
&\ket{\psi(2m\tau)} = \bigl(\hat{U}_F\bigr)^m \ket{\psi(0)}, \\
&\hat{U}_F = \hat{\mathcal{T}} \exp \biggl( i \int_0^{2\tau} dt\, \Delta_p(t) \hat{N}_{\mathrm{I}}(t) \biggr).
\end{split}
\end{equation}
We may now perform a high frequency expansion of the Floquet system defined by \eq{eq:app2}, provided that the frequency of the periodic drive is fast compared to the natural time scale of the detuning perturbation, $\|\Delta_p \| \ll 1/\tau$.
To leading order in $\Delta_p(t)$, $\hat{U}_F \approx \exp(-i 2\tau \hat{\mathcal{H}})$, with the effective Hamiltonian
\begin{equation} \label{eq:app3}
\hat{\mathcal{H}} = -\frac{1}{2\tau}\int_0^{2\tau} dt\, \Delta_p(t) \hat{N}_{\mathrm{I}}(t),
\end{equation}
given by the time average of $\Delta_p(t)\hat{N}_{\mathrm{I}}(t)$ over one period of the drive.
Using the form of $\hat{U}_e(t)$ in \eq{eq:app1}, the number operator in the interaction picture is given by
\begin{equation} \label{eq:app4}
\begin{split}
&\hat{N}_{\mathrm{I}}(0\leq t < \tau) = e^{it \hat{H}_0} \hat{N} e^{-it \hat{H}_0} = \hat{N}_0(t) \\
&\hat{N}_{\mathrm{I}}(\tau \leq t \leq 2\tau) = \hat{N}_0(2\tau-t).
\end{split}
\end{equation}
Inserting into \eq{eq:app3} and using that $\int_\tau^{2\tau} dt \, \Delta_p(t) \hat{N}_0(2\tau-t) = \int_0^{\tau} dt \, \Delta_p(2\tau-t) \hat{N}_0(t)$, we obtain
\begin{equation}
\hat{\mathcal{H}} = -\int_{0}^{2\tau} \frac{dt}{2\tau} \Bigl[ \Delta_p(t) + \Delta_p(2\tau-t) \Bigr] \hat{N}_0(t).
\end{equation}
We thus see that $\hat{\mathcal{H}}$ picks up contributions proportional to $\hat{N}_0(t)$ from both forward and backward evolution of the many-body echo. Finally, assuming symmetry of $\Delta_p(t)$ around $t=\tau$ leads to \eq{eq:2.2.2} of the main text.

We note that according to Ref.~\cite{else2017_dtc}, the prethermal timescale $t_*$ on which dynamics is approximately described by a local Hamiltonian is bounded by $t_* \gtrsim \exp\{C / \int_0^\tau dt |\Delta_p(t)|\}$, i.e., by the one-norm of $\Delta_p(t)$ ($C>0$ is an $\mathcal{O}(1)$ constant). However, using Cauchy-Schwarz, $\int_0^\tau dt |\Delta_p(t)| \leq \bigl( \int_0^\tau dt \tau |\Delta_p(t)|^2 \bigr)^{1/2}$, and we may thus also use the two-norm of $\Delta_p(t)$ to bound $t_*$. While this provides a less optimal bound, the two-norm is a convenient ingredient in the optimization scheme for the detuning profile described in Sec.~\ref{sec:opt1}.
Nonetheless, the one-norm in principle also allows for \textit{discrete} perturbations around the many-body echo, i.e., detuning perturbations of the form
\begin{equation}
\Delta_d(t) = \sum_{j} \frac{\Delta_j}{\Omega} \delta(t-t_j).
\end{equation}
The corresponding contribution $\hat{\mathcal{H}}_d$ to the effective Hamiltonian is given by 
\begin{equation}
\hat{\mathcal{H}}_d = - \sum_{0\leq t_j < \tau} \frac{\Delta_j}{2\tau \Omega} \hat{N}_0(t_j) - \sum_{\tau\leq t_j \leq 2\tau} \frac{\Delta_j}{2\tau \Omega} \hat{N}_0(2\tau - t_j).
\end{equation}
Timed pulses thus allow to extract contributions of $\hat{N}_0(t_j)$ at specific instances $t_j$. 
In this work, we only use this possibility for adding a discrete pulse of strength $\Delta_0$ at times $t=2\tau m$, which accordingly enters the effective Hamiltonian $\hat{\mathcal{H}}$ as a static detuning field, $\hat{\mathcal{H}}_d = -\frac{\Delta_0}{2\tau\Omega}\hat{N}$. We note that this corresponds to changing the weight of the $\pi$-pulses that are already applied at times $2\tau m$ as part of the echo protocol, \eq{eq:app_echo}.

\begin{figure*}[t]
\centering
\includegraphics[trim={0cm 0cm 0cm 0cm},clip,width=0.99\linewidth]{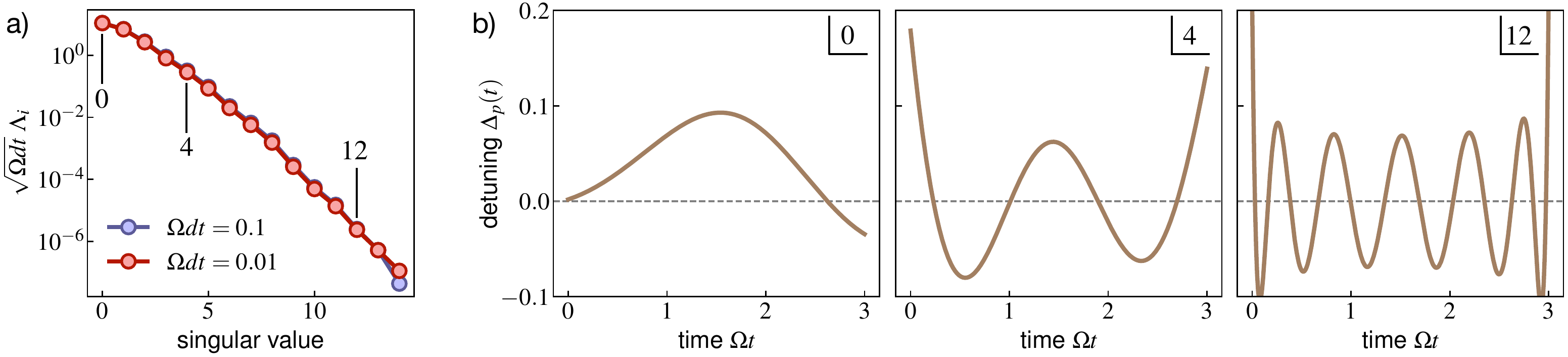}
\caption{\textbf{Singular value decomposition.} \textbf{a)} The 14 largest singular values $D_i$ of the matrix $\underline{N_0}$ (see text) for a period $\Omega\tau = 3.0$, evaluated for time step $\Omega dt=0.1$ and $\Omega dt = 0.01$. The singular values scale as $D_i\sim 1/\sqrt{dt}$. \textbf{b)} Selected detuning profiles $\Delta_p(t)$ associated to the singular values in a). As the magnitude of singular values becomes smaller, $\Delta_p(t)$ oscillates more rapidly in time. 
}
\label{fig:singular_values}
\end{figure*}

\section{Time-evolved operators: Matrix elements} \label{app:operator_basis}

We consider the time-evolved operator $\hat{N}_0(t)$ entering the expression for the effective Hamiltonian in \eq{eq:2.2.2}. Let $\ket{n}$, $\ket{n^\prime}$ be two occupation number product states containing $N_n$ and $N_{n^\prime}$ excitations, respectively. Using the anticommutation $\{\prod_i \hat{\sigma}^z_i,\hat{H}_0\}=0$ of the parity operator with the PXP Hamiltonian $\hat{H}_0$, we obtain
\begin{equation} \label{eq:app3.7}
\begin{split}
&\braket{n^\prime|\hat{N}_0(t)|n} = \\
& = \braket{n^\prime|\biggl(\prod_i \hat{\sigma}^z_i\biggr)^2 e^{i\hat{H}_0 t} \hat{N} e^{-i\hat{H}_0 t} \biggl(\prod_i \hat{\sigma}^z_i\biggr)^2|n} = \\
&=(-1)^{N_n+N_{n^\prime}} \braket{n^\prime|\biggl(\prod_i \hat{\sigma}^z_i\biggr) e^{i\hat{H}_0 t} \hat{N} e^{-i\hat{H}_0 t} \biggl(\prod_i \hat{\sigma}^z_i\biggr)|n} = \\
&=(-1)^{N_n+N_{n^\prime}} \braket{n^\prime| e^{-i\hat{H}_0 t} \hat{N} e^{i\hat{H}_0 t} |n} = \\
&= (-1)^{N_n+N_{n^\prime}} \braket{n^\prime|\hat{N}_0(-t)|n} = \\
&= (-1)^{N_n+N_{n^\prime}} \bigl(\braket{n^\prime|\hat{N}_0(t)|n}\bigr)^*.
\end{split}
\end{equation}
The last equality in \eq{eq:app3.7} follows from the fact that the matrix elements of $\hat{H}_0$ are real.
Due to \eq{eq:app3.7}, matrix elements of the (leading order) effective Hamiltonian between basis states whose Rydberg excitation numbers differ by an even/odd integer are purely real/imaginary. 
Furthermore, due to Rydberg blockade, for nearest neighbor sites $i,j$ we have
\begin{equation} \label{eq:app3.4}
\hat{P} \hat{\sigma}^x_i \hat{\sigma}^x_j \hat{P} = \hat{P} \hat{\sigma}^y_i \hat{\sigma}^y_j \hat{P}.
\end{equation}
Consequently, offdiagonal contributions to the effective Hamiltonian in which spins are flipped along \textit{connected} paths $\Lambda$ of length $|\Lambda| = \gamma$ can be written purely in terms of $\hat{\sigma}^y_i$ operators. This leads to the operators $\hat{O}_\gamma$ of \eq{eq:offdiagonal_ops}.

Using the result of \eq{eq:app3.7}, we can now show that the matrix $\underline{N_0}^\dagger\underline{N_0}$ used in the optimization of \eq{eq:2.3.3} is real (and thus symmetric, since hermiticity is manifest). Specifically,
\begin{equation} \label{eq:matrix_real_and_symm}
\begin{split}
&(\underline{N_0}^\dagger\underline{N_0})_{i,j} = \mathrm{Tr}[\hat{N}^\dagger_0(t_i) \hat{N}_0(t_j)] = \\
&= \sum_{n,n^\prime} \braket{n|\hat{N}^\dagger_0(t_i)|n^\prime} \braket{n^\prime|\hat{N}_0(t_j)|n} = \\
&= \sum_{n,n^\prime} \bigl(\braket{n^\prime|\hat{N}_0(t_i)|n}\bigr)^* \braket{n^\prime|\hat{N}_0(t_j)|n} = \\
&=  \sum_{n,n^\prime} \braket{n^\prime|\hat{N}_0(t_i)|n} \bigl( \braket{n^\prime|\hat{N}_0(t_j)|n} \bigr)^* = (\underline{N_0}^\dagger\underline{N_0})_{i,j}^*.
\end{split}
\end{equation}
To go from the third to the fourth line in \eq{eq:matrix_real_and_symm} we have used \eq{eq:app3.7} twice.
Moreover, since the eigenvalues of $\underline{N_0}^\dagger\underline{N_0}$ are given by the squared magnitudes of the singular values of $\underline{N_0}$, it follows that $\underline{N_0}^\dagger\underline{N_0}$ is positive semidefinite.
As $\underline{N_0}^\dagger\underline{N_0}$ is real and symmetric, its eigenvectors can always be chosen real. Equivalently, the matrix $V$ in the singular value decomposition $\underline{N_0}=W\Lambda V^\dagger$ (whose columns correspond to discretized detuning profiles) can be chosen real.

\section{Time-evolved operators: Singular value decomposition} \label{app:operator_SVD}
In this appendix, we provide additional details on the singular value decomposition $\underline{N_0}=WD V^\dagger$ of the matrix $\underline{N_0}=(\bs{N}_0(t_1),...,\bs{N}_0(t_M)$, where $t_1=0$, $t_M=\tau$. For concreteness, we fix a period $\Omega\tau=3.0$ and a subspace of relevant matrix elements as in Sec.~\ref{sec:three_body},
\begin{equation} \label{eq:app_matrix_elements}
S = \bigl\{ (\ket{n_{k=0}},\ket{n^\prime_{k=0}}) : N_n, N_{n^\prime} \geq L^2-1 \bigr\}.
\end{equation}
$\underline{N_0}$ is thus a $|S|\times M$ matrix, where $M=\tau/dt$ is inversely proportional to the time step $dt = t_{i+1}-t_i$.
We subsequently evaluate the SVD for a discretized time step $\Omega dt = 0.1$, as well as for $\Omega dt = 0.01$. The number of columns in $\underline{N_0}$ increases linearly in $1/dt$, and the resulting singular values thus scale as $D_i\sim 1/\sqrt{dt}$, as confirmed numerically in \figc{fig:singular_values}{a}.
In \figc{fig:singular_values}{b} we display the detuning profiles, i.e., the columns of the matrix V, associated with the corresponding singular values. We observe that as the magnitude $D_i$ of the singular values decreases, the associated profiles exhibit increasingly rapid oscillations. This is in accordance with our physical expectation: The time-evolved operator $\hat{N}_0(t)$ varies on a characteristic time scale set by the Rabi frequency $\Omega$ of the underlying PXP model. Detuning profiles oscillating at frequencies much higher than $\Omega$ predominantly lead to destructive interference, resulting in a small prefactor (and thus singular value) of the engineered interaction term.
As such, a highly precise implementation of Hamiltonians that require large amounts of descructive interference is difficult, while `easy' Hamiltonians rely primarily on constructive interference.

\begin{figure*}[t]
\centering
\includegraphics[trim={0cm 0cm 0cm 0cm},clip,width=0.99\linewidth]{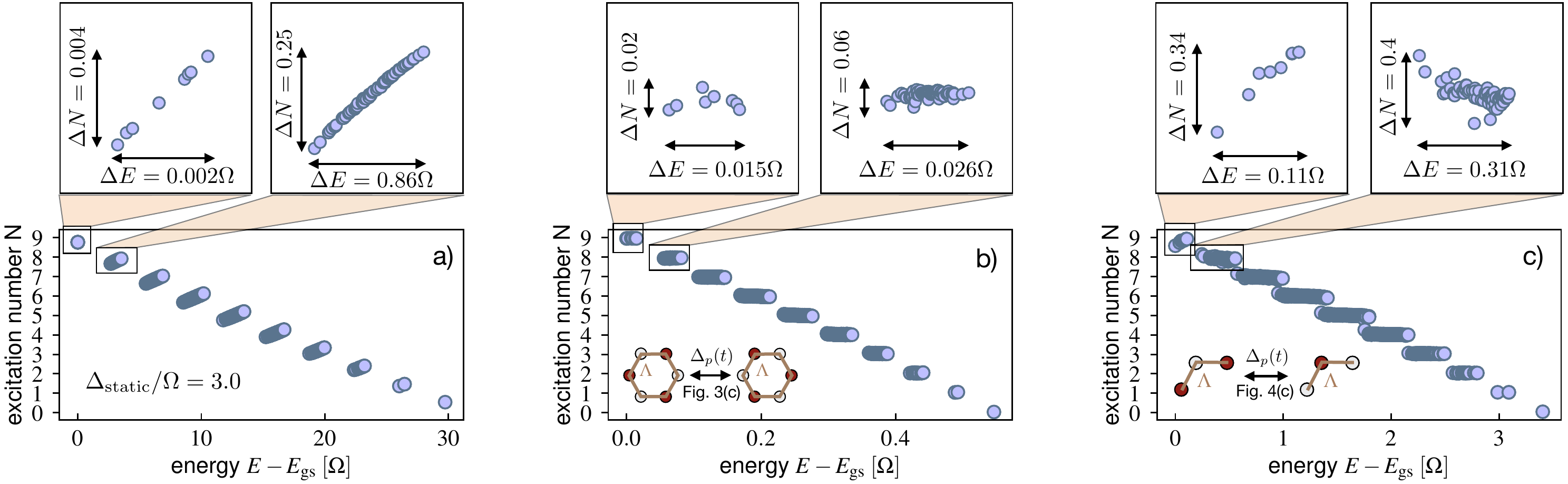}
\caption{\textbf{Eigenspectrum of effective Hamiltonians.} \textbf{a)} Energies and global Rydberg number expectation values of the $k=0$ momentum eigenstates of a PXP model at static detuning $\Delta_{\mathrm{static}}/\Omega=3.0$. Energy bands with approximate number conservation become apparent. There is a large separation of scales in the bandwidths of the fully packed sector and the remainder of the spectrum. The tilt of the bands in the $E-N$ plane is due to the perturbative nature of the processes that connect different basis states with the same global occupation number. \textbf{b)} Same as a) for the effective Hamiltonian of Sec.~\ref{sec:six_body} targeting six-body terms. The bandwidth of the fully packed sector is now comparable to other sectors. Due to the dominant multi-body interactions in $\hat{\mathcal{H}}$ directly connecting states within a global number sector, energy bands are not tilted in the $E-V$ plane. \textbf{c)} Spectrum for the effective Hamiltonian of Sec.~\ref{sec:three_body}, with the continuous profile $\Delta_p(t)$ rescaled by a factor $0.7$ relative to \figc{fig:f2_opt}{c}. Only the fully packed sector exhibits a tilted energy band, indicating dominant perturbative processes only for pure gauge dynamics.}
\label{fig:spectral_properties}
\end{figure*}

\section{Spectral properties of effective Hamiltonian}
In the main text, we have explored different avenues to verify that the dynamics of the effective Hamiltonian (as well as the stroboscopic Floquet evolution) is dominated by the engineered interaction terms. This includes dynamics under rescaling of the offdiagonal part of $\hat{\mathcal{H}}$ as well as analyzing occupation number snapshots from specific initial states. Here, we show that related insights can be gained from analyzing the spectrum of $\hat{\mathcal{H}}$.

We first consider the eigenstates in the $k=0$ momentum sector of the static PXP Hamiltonian at constant detuning $\Delta/\Omega = 3$ in \figc{fig:spectral_properties}{a}. We resolve the eigenstates with respect to their energy and global Rydberg number expectation value. As expected, the spectrum decomposes into sectors of approximately constant occupation numbers $N$. In addition, the energy bandwidths $\Delta E_{N}$ of the individual number sectors are inverse related to their characteristic timescale of dynamics. In particular, zooming in on the $N=L^2$ and $N=L^2-1$ sectors, the large separation of timescales between gauge fields and charge excitations, previously observed in real time dynamics, translates into $\Delta E_{L^2} \ll \Delta E_{L^2-1}$.
Finally, we observe that the bands are tilted in the $E-N$ plane of eigenstate energy and number expectation value. This tilt is a direct consequence of the fact that dynamics within a given number sector is induced \textit{perturbatively} in $\Omega/\Delta$ via processes that change the global occupation number. Formally, the effective Hamiltonian of \eq{eq:Hstat} at large static detuning is valid in a dressed basis, $\hat{H} = \hat{U}_{SW} \hat{H}_{\mathrm{static}} \hat{U}_{SW}^\dagger$, where $\hat{U}_{SW}$ is a Schrieffer-Wolff basis transformation that is perturbative in $\Omega/\Delta$ and number-changing.

In comparison, \figc{fig:spectral_properties}{b} shows the spectrum of the effective Hamiltonian associated to the protocol of Sec~\ref{sec:six_body} targeting six-body terms. Again, we find clearly separated, approximately number-conserving sectors. The bandwidths $\Delta E_{L^2} \lesssim \Delta E_{L^2-1}$ are of more similar strengths. The most striking difference to \figc{fig:spectral_properties}{a} is the absence of a tilt of the bands in the $E-N$ plane. This is indicative of direct matrix elements (instead of perturbatively generated processes) that lift the degeneracy between states of equal global occupation number. 
For the protocol of Sec~\ref{sec:three_body} that targets three-body interactions, see \figc{fig:spectral_properties}{c}, a tilt in the $E-N$ plane is present only for the fully packed sector with $N=L^2$ excitations. This agrees with our expectation that only pure gauge dynamics requires a perturbative process.

\section{Multi-body dynamics in snapshot sampling} \label{sec:snapshots}
\begin{figure}[t]
\centering
\includegraphics[trim={0cm 0cm 0cm 0cm},clip,width=0.99\linewidth]{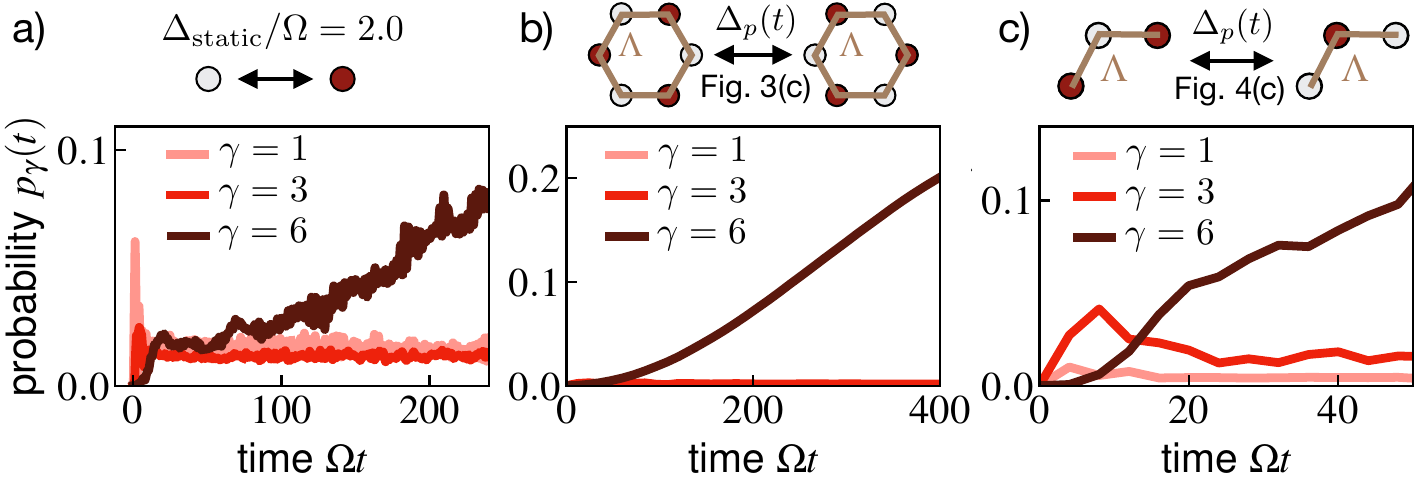}
\caption{\textbf{Transition graph dynamics.} To verify the mechanism behind the pure gauge dynamics, we evaluate the probabilities $p_\gamma(t)$ of finding connected paths of length $\gamma$ along which spins are flipped relative to the dimer initial state $\ket{\psi_1}$, see \eq{eq:path_snap}. The static protocol a) shows a significant value of $p_1(t)$ at early times, consistent with single-body spin flips. In b), all $p_{\gamma < 6}(t)$ remain close to zero, indicating dominant six-body plaquette terms. In c), $p_{3}(t) \gg p_1(t)$ is large at early times, consistent with dominant three-body terms.  }
\label{fig:transition_graphs}
\end{figure}

In the main text, we have verified the presence of strong plaquette terms in our Floquet protocols.
In addition, we want to infer from local observables the mechanism by which they are engineered. 
For this purpose, we consider snapshots of a time-evolved dimer initial state $\ket{\psi(t)}$ in the occupation number basis. To quantify the degree to which the resulting snapshots differ from the initial $\ket{\psi}$ by application of specific $\gamma$-body offdiagonal operators, we consider the probabilities
\begin{equation}	\label{eq:path_snap}
\begin{split}
&p_\gamma(t) = \\
&\frac{1}{3L^2} \sum_{n^\prime} \left|\braket{n^\prime|\psi(t)}\right|^2 \, \sum_i \delta \bigl( \gamma - \max_{\Lambda:\, i \in \Lambda } \, |\Lambda| \prod_{j \in \Lambda} \bigl| n^\prime_j - \braket{\hat{n}_j}_\psi \bigr| \bigr)
\end{split}
\end{equation}
This quantity has a simple interpretation: The first sum samples occupation number basis states from the time-evolved state $\ket{\psi(t)}$. We then examine the `transition graph' between a given outcome $\ket{n^\prime}$ and the initial state $\ket{\psi}$, i.e., the sites at which the two states differ from each other. Each such site contributes to $p_\gamma(t)$ if it is part of a connected path of differing sites of length $\gamma$. We thus have $\sum_{\gamma = 0}^{3L^2} p_\gamma (t) = 1$. We show $p_\gamma(t)$ starting from the dimer state $\ket{\psi_1}$ in \fig{fig:transition_graphs}. 
Common to all protocols is the growth of $p_6(t)$ due to the action of plaquette terms. However, for the static case, \figc{fig:transition_graphs}{a}, the probability $p_1(t)$ is large at early times and retains a significant value throughout the shown times. In contrast, for the six-body Floquet protocol of Sec.~\ref{sec:six_body}, the probabilites $p_{\gamma < 6}(t)$ remain very close to zero. For the three-body Floquet scheme of Sec.~\ref{sec:three_body}, an initial peak in $p_3(t)$ occurs, and $p_3(t) \gg p_1(t)$ for the plotted times. These results are consistent with single-body, six-body, and three-body offdiagonal operators being the dominant generators of dynamics in each of the protocols, respectively. We note that the corresponding detuning profiles of \figc{fig:f1_opt}{c} and \figc{fig:f2_opt}{c} were optimized specifically in the $k=0$ momentum sector, but are seen to generalize well to other sectors here. The approach of verifying the terms of an effective Hamiltonian via occupation basis snapshots is readily accessible in current quantum simulation hardware.  

\section{Properties of the optimization procedure} \label{sec:optimization}
In this appendix we provide additional details for the optimization scheme described in Sec.~\ref{sec:method} that determines the detuning perturbation $\Delta_p(t)$.

\subsection{Subspace-restricted optimization} \label{sec:subspaces}
As stated in the main text, we often optimize the effective Hamiltonian only with respect to a subset of all possible matrix elements. Formally, we denote by
\begin{equation}
S = \{ (\ket{s_i},\ket{s^\prime_i}): i=1,...,|S| \}
\end{equation}
a set of pairs of states $(\ket{s_i},\ket{s^\prime_i})$, taken from an orthogonal basis of the Hilbert space. Within the set of matrix elements determined by $S$, we define an inner product and norm for operators as
\begin{equation} \label{eq:inner_product}
(\hat{O}_1,\hat{O}_2) \equiv \sum_{i=1}^{|S|} \braket{s_i|\hat{O}_1|s^\prime_i}^* \braket{s_i|\hat{O}_2|s^\prime_i},
\end{equation}
\begin{equation}
\| \hat{O} \|^2 \equiv (\hat{O},\hat{O}) = \sum_{i=1}^{|S|} |\braket{s_i|\hat{O}|s^\prime_i}|^2.
\end{equation}

\subsection{Subtracting traceful contributions}
We may generally allow for an arbitrary traceful contribution $\sim \mathbb{1}$ to the effective Hamiltonian $\hat{\mathcal{H}}$, since this merely results in a global dynamical phase. We therefore optimize only the traceless part of $\hat{\mathcal{H}}$ in \eq{eq:2.3.2}. This can be achieved by subtracting the trace-part of both the target operator, $\hat{T} \rightarrow \hat{T} - \frac{\mathrm{Tr}[\hat{T}]}{\mathrm{Tr}[\mathbb{1}]} \mathbb{1}$, as well as the time evolved number operator, $\hat{N}_0(t) \rightarrow \hat{N}_0(t) - \frac{\mathrm{Tr}[\hat{N}]}{\mathrm{Tr}[\mathbb{1}]} \mathbb{1}$.

\subsection{Continuity of the detuning profile}
The (discretized) detuning solution $\bs{\tilde{\Delta}}_p$ of \eq{eq:2.3.3} of the main text that minimizes the cost function \eq{eq:2.3.2} inherits its continuity (in time) from the manifest smoothness of the time-evolved $\hat{N}_0(t)$ when $\lambda > 0$. Formally, as $dt\rightarrow 0$,
\begin{equation} \label{eq:app2.1}
\lim_{dt\rightarrow 0}\,  \biggl[ \tilde{\Delta}_p(t_{i+1}) -  \tilde{\Delta}_p(t_{i}) \biggr] = 0,
\end{equation}
where $t_{i+1}=t_i+dt$.

In order to show this, let us consider the explicit dependence of the cost function \eq{eq:2.3.2} on the discretized detuning profile at the two time slices $t_{i+1}$ and $t_i$: 
\begin{equation} \label{eq:app2.2}
\begin{split}
&C_{\lambda,\tau}(\bs{T}) = \\ 
&=\| \bs{T}^\prime + \frac{dt}{\tau} \bs{N}_0(t_i) \Delta_p(t_i) + \frac{dt}{\tau} \bs{N}_0(t_{i+1}) \Delta_p(t_{i+1}) \|^2 \\
&+ dt \lambda \|\bs{\Delta}^\prime_p \|^2 + dt \lambda \bigl[\Delta_p(t_i)^2 + \Delta_p(t_{i+1})^2 \bigr],
\end{split}
\end{equation}
where $\bs{T}^\prime \equiv \bs{T} - \bs{\mathcal{H}}[\bs{\Delta}_p] - \frac{dt}{\tau} \bs{N}_0(t_i) \Delta_p(t_i) - \frac{dt}{\tau} \bs{N}_0(t_{i+1}) \Delta_p(t_{i+1})$ and $\|\bs{\Delta}^\prime_p \|^2 \equiv \|\bs{\Delta}_p \|^2 - \Delta_p(t_i)^2 - \Delta_p(t_{i+1})^2$ are independent of $\Delta_p(t_i)$, $\Delta_p(t_{i+1})$. Using that 
\begin{equation}
\begin{split}
\bs{N}_0(t_i) = \bs{N}_0(t_i+dt/2) - \frac{dt}{2} (\partial_t \bs{N}_0)(t_i+dt/2) \\ 
\bs{N}_0(t_{i+1}) = \bs{N}_0(t_i+dt/2) + \frac{dt}{2} (\partial_t \bs{N}_0)(t_i+dt/2),
\end{split}
\end{equation}
\eq{eq:app2.2} can be written as
\begin{equation} \label{eq:app2.3}
\begin{split}
&C_{\lambda,\tau}(\bs{T}) = \\ 
&=\bigl\| \bs{T}^\prime + \frac{dt}{\tau} \bs{N}_0(t_i+dt/2) \bigl[ \Delta_p(t_i) + \Delta_p(t_{i+1}) \bigr] + \\
& \qquad \qquad + \frac{dt^2}{2\tau}(\partial_t \bs{N}_0)(t_i+dt/2)\bigl[ \Delta_p(t_{i+1}) - \Delta_p(t_{i}) \bigr]\, \bigr\|^2 + \\
& \quad +dt \lambda \|\bs{\Delta}^\prime_p \|^2 + \frac{dt}{2} \lambda \bigl[ \Delta_p(t_i) + \Delta_p(t_{i+1}) \bigr]^2 + \\
& \quad + \frac{dt}{2} \lambda \bigl[ \Delta_p(t_{i+1}) - \Delta_p(t_{i}) \bigr]^2.
\end{split}
\end{equation}
\eq{eq:app2.3} expresses $C_{\lambda,\tau}(\bs{T})$ in terms of the symmetric and antisymmetric combinations of $\Delta_p(t_i)$ and $\Delta_p(t_{i+1})$. In order to minimize the cost function, we now perform the derivative of \eq{eq:app2.3} with respect to the antisymmetric combination $\Delta_p(t_{i+1})-\Delta_p(t_{i})$ and demand $\partial_{\Delta_p(t_{i+1})-\Delta_p(t_{i})}C_{\lambda,\tau}(\bs{T})=0$. This condition can be solved for $\Delta_p(t_{i+1})-\Delta_p(t_{i})$, resulting in
\begin{equation} \label{eq:app2.4}
\begin{split}
&\tilde{\Delta}_p(t_{i+1})-\tilde{\Delta}_p(t_{i}) = \\ 
&-\frac{dt^2}{2\tau}\frac{\mathrm{Re}\biggl[ \bs{T}^{\prime\prime} \cdot (\partial_t \bs{N}_0)^*(t_i+dt/2) \biggr]}{\lambda dt/2 + dt^4/(4\tau^2)\, \|(\partial_t \bs{N}_0)(t_i+dt/2)\|^2},
\end{split}
\end{equation}
with $\bs{T}^{\prime\prime} = \bs{T}^\prime + \frac{dt}{\tau} \bs{N}_0(t_i+dt/2) \bigl[ \Delta_p(t_i) + \Delta_p(t_{i+1}) \bigr]$.
Collecting factors of $dt$ and using $\lambda >0$, we see from \eq{eq:app2.4} that
\begin{equation}
\tilde{\Delta}_p(t_{i+1})-\tilde{\Delta}_p(t_{i}) = \mathcal{O}(dt)  \xrightarrow{dt\rightarrow 0} 0,
\end{equation}
which proves \eq{eq:app2.1}.

\subsection{Proof of $\partial_\lambda Q_\Delta(\lambda,\tau) \geq 0$}

We start with the solution $\bs{\tilde{\Delta}}_p$ in \eq{eq:2.3.3}. For simplicity, we assume that the target vector $\bs{T}$ is normalized to unity. 
According to \eq{eq:2.3.1}, the $\bs{\mathcal{H}}$ resulting from $\bs{\tilde{\Delta}}_p$ is
\begin{equation}
\bs{\mathcal{H}} = -\frac{dt}{\tau}\underline{N_0}\cdot \bs{\tilde{\Delta}}_p = \frac{dt}{\tau} \underline{N_0}\, \biggl( \frac{dt}{\tau}\underline{N_0}^\dagger \underline{N_0} + \tau \lambda \mathbb{1} \biggr)^{-1} \underline{N_0}^\dagger \cdot \bs{T}. 
\end{equation}
Its projection along the target direction is given by 
\begin{equation} \label{eq:app2.6}
\begin{split}
\bs{T}^* \cdot \bs{\mathcal{H}} = \frac{dt}{\tau} \bs{T}^* \cdot \underline{N_0}\, \biggl( \frac{dt}{\tau}\underline{N_0}^\dagger \underline{N_0} + \tau \lambda \mathbb{1} \biggr)^{-1} \underline{N_0}^\dagger \cdot \bs{T}.
\end{split}
\end{equation}
We now diagonalize the positive semidefinite $M \times M$ matrix $\frac{dt}{\tau}\underline{N_0}^\dagger \underline{N_0} = V\, \Gamma\, V^\dagger$. Here, $\Gamma$ is a diagonal matrix containing the eigenvalues $\Gamma_i = \frac{dt}{\tau} \Lambda_i^2\geq 0$, where $\Lambda_i$ are the singular values of $\underline{N_0}$. Using this decomposition and defining the vector $\bs{p} \equiv V^\dagger \underline{N_0}^\dagger\cdot \bs{T}$, \eq{eq:app2.6} reads
\begin{equation} \label{eq:app2.7}
\begin{split}
\bs{T}^* \cdot \bs{\mathcal{H}} = \frac{dt}{\tau} \sum_i \frac{|p_i|^2}{\Gamma_i+\tau\lambda}.
\end{split}
\end{equation}
At the same time, the norm of the optimized detuning profile $\bs{\tilde{\Delta}}_p$ is given by
\begin{equation} \label{eq:app2.8}
\begin{split}
\|\bs{\tilde{\Delta}}_p\|^2 &= \bs{T}^* \cdot \underline{N_0}\, \biggl( \frac{dt}{\tau}\underline{N_0}^\dagger \underline{N_0} + \tau\lambda \mathbb{1} \biggr)^{-2} \underline{N_0}^\dagger \cdot \bs{T} = \\
&= \sum_i \frac{|p_i|^2}{\bigl(\Gamma_i + \tau\lambda\bigr)^2}.
\end{split}
\end{equation}
Equipped with \eq{eq:app2.7} and \eq{eq:app2.8}, we compute 
\begin{equation} \label{eq:app2.9}
\begin{split}
&\partial_\lambda Q_\Delta(\lambda,\tau) = \partial_\lambda \biggl[\frac{\bs{T}^*\cdot \bs{\mathcal{H}}}{\|\bs{\tilde{\Delta}}_p\|} \biggr] \propto \\
&\propto \partial_\lambda \bigl[\bs{T}^*\cdot \bs{\mathcal{H}}\bigr]\, \|\bs{\tilde{\Delta}}_p\| - \bigl(\bs{T}^*\cdot \bs{\mathcal{H}}\bigr) \, \partial_\lambda \|\bs{\tilde{\Delta}}_p\|= \\
% an in between step here
%= dt \biggl( \sum_i \frac{-|p_i|^2}{\bigr(\frac{dt}{\tau}S_i + \tau \lambda \bigr)^2} \biggr) \biggl( \sum_i \frac{|p_i|^2}{\bigl(\frac{dt}{\tau}S_i+\tau\lambda \bigr)^2} \biggr)^{1/2} + \\
%+ dt \biggl( \sum_i \frac{|p_i|^2}{\frac{dt}{\tau}S_i + \tau \lambda} \biggr) \biggl( \sum_i \frac{|p_i|^2}{\bigr(\frac{dt}{\tau}S_i + \tau \lambda \bigr)^2} \biggr)^{-1/2} \biggl( \sum_i \frac{|p_i|^2}{\bigr(\frac{dt}{\tau}S_i + \tau \lambda \bigr)^3} \biggr) = \\
&= dt \biggl( \sum_i \frac{|p_i|^2}{\bigr(\Gamma_i + \tau \lambda \bigr)^2} \biggr)^{-1/2} \sum_{i,j} \frac{|p_i|^2 |p_j|^2 \bigl( \Gamma_i - \Gamma_j \bigr) }{\bigl( \Gamma_i+\tau\lambda\bigr)^2 \bigl( \Gamma_j+\tau\lambda\bigr)^3}.
\end{split}
\end{equation}
The term in \eq{eq:app2.9} in the double sum can be rewritten as
\begin{equation} \label{eq:app2.10}
\begin{split}
&\sum_{i,j} \frac{|p_i|^2 |p_j|^2 \bigl( \Gamma_i - \Gamma_j \bigr) }{\bigl( \Gamma_i+\tau\lambda\bigr)^2 \bigl( \Gamma_j+\tau\lambda\bigr)^3} = \\
&=\sum_{i,j: \Gamma_i>\Gamma_j} \frac{|p_i|^2 |p_j|^2 \bigl( \Gamma_i - \Gamma_j \bigr) }{\bigl( \Gamma_i+\tau\lambda\bigr)^2 \bigl( \Gamma_j+\tau\lambda\bigr)^2}\Biggl[ \frac{1}{\Gamma_j  +\tau\lambda} - \frac{1}{\Gamma_i + \tau\lambda}\Biggr].
\end{split}
\end{equation}
As $\Gamma_i>\Gamma_j\geq 0$ in \eq{eq:app2.10} and $\lambda > 0$, the term in square brackets on the right hand side of \eq{eq:app2.10} is positive. Therefore, the full expression in \eq{eq:app2.10} is non-negative, which proves $\partial_\lambda Q_\Delta(\lambda,\tau)\geq 0$.

\subsection{Proof of $\partial_\lambda Q_{\measuredangle}(\lambda,\tau) \leq 0$}

This relation can be shown similarly to the previous one by diagonalizing the matrix $\frac{dt}{\tau}\underline{N_0}^\dagger \underline{N_0}$. In addition to the expression of \eq{eq:app2.7}, we will need the norm of $\bs{\mathcal{H}}$, which is readily calculated as
\begin{equation} \label{eq:app2.11}
\|\bs{\mathcal{H}}\|^2 = \frac{dt}{\tau} \sum_i |p_i|^2 \frac{\Gamma_i}{\bigl(\Gamma_i +  \tau\lambda\bigr)^2}.
\end{equation}
We can now evaluate 
\begin{equation} \label{eq:app2.12}
\begin{split}
&\partial_\lambda \biggl[ \frac{|\bs{T}^*\cdot\bs{\mathcal{H}}|^2}{\|\bs{\mathcal{H}}\|^2} \biggr] \propto \\
&\propto \partial_\lambda \bigl[ |\bs{T}^*\cdot\bs{\mathcal{H}}|^2 \bigr] \; \|\bs{\mathcal{H}}\|^2 - |\bs{T}^*\cdot\bs{\mathcal{H}}|^2 \; \partial_\lambda \bigl[ \|\bs{\mathcal{H}}\|^2 \bigr] = \\
&= \frac{dt^3}{\tau^2}\sum_{i,j,k} \frac{|p_i|^2|p_j|^2|p_k|^2}{\bigl(\Gamma_i+\tau\lambda \bigr) \bigl(\Gamma_j+\tau\lambda \bigr) \bigl(\Gamma_k+\tau\lambda \bigr)} \times \\
&\qquad \times \frac{\Gamma_i}{\Gamma_i+\tau\lambda}\biggl[ \frac{2}{\Gamma_i+\tau\lambda} - \frac{1}{\Gamma_j+\tau\lambda} - \frac{1}{\Gamma_k+\tau\lambda} \biggr],
\end{split}
\end{equation}
which follows after inserting \eq{eq:app2.7}, \eq{eq:app2.11} and a few lines of algebra. Due to the triple sum over $i,j,k$ in \eq{eq:app2.12}, we can symmetrize the expression inside the sum over all permutations of $i,j,k$. \eq{eq:app2.12} then becomes
\begingroup
\begin{align} \label{eq:app2.13}
%\begin{split}
&\partial_\lambda \biggl[ \frac{|\bs{T}^*\cdot\bs{\mathcal{H}}|^2}{\|\bs{\mathcal{H}}\|^2} \biggr] \propto \nonumber \\
&\propto \frac{dt^3}{\tau^2}\sum_{i,j,k} \frac{|p_i|^2|p_j|^2|p_k|^2}{\bigl(\Gamma_i+\tau\lambda \bigr) \bigl(\Gamma_j+\tau\lambda \bigr) \bigl(\Gamma_k+\tau\lambda \bigr)} \times \nonumber \\
&\times \Biggl\{ \frac{\Gamma_i}{\Gamma_i+\tau\lambda}\biggl[ \frac{2}{\Gamma_i+\tau\lambda} - \frac{1}{\Gamma_j+\tau\lambda} - \frac{1}{\Gamma_k+\tau\lambda} \biggr] + \\
&+ \frac{\Gamma_j}{\Gamma_j+\tau\lambda}\biggl[ \frac{2}{\Gamma_j+\tau\lambda} - \frac{1}{\Gamma_i+\tau\lambda} - \frac{1}{\Gamma_k+\tau\lambda} \biggr] + \nonumber \\
&+ \frac{\Gamma_k}{\Gamma_k+\tau\lambda}\biggl[ \frac{2}{\Gamma_k+\tau\lambda} - \frac{1}{\Gamma_j+\tau\lambda} - \frac{1}{\Gamma_i+\tau\lambda} \biggr] \Biggr\}. \nonumber
%\end{split}
\end{align}
\endgroup
The term in the large curly bracket in \eq{eq:app2.13} can be rearranged as
\begin{equation} \label{eq:app2.14}
\begin{split}
\Biggl\{\cdot\cdot\cdot \Biggr\} = 
&\biggl[ \frac{1}{\Gamma_i+\tau\lambda} - \frac{1}{\Gamma_j+\tau\lambda} \biggr]\, \biggl[ \frac{\tau\lambda \bigl(\Gamma_i - \Gamma_j\bigr)}{\bigl(\Gamma_i+\tau\lambda \bigr) \bigl(\Gamma_j+\tau\lambda \bigr)} \biggr] + \\
+ &\biggl[ \frac{1}{\Gamma_i+\tau\lambda} - \frac{1}{\Gamma_k+\tau\lambda} \biggr]\, \biggl[ \frac{\tau\lambda \bigl(\Gamma_i - \Gamma_k\bigr)}{\bigl(\Gamma_i+\tau\lambda \bigr) \bigl(\Gamma_k+\tau\lambda \bigr)} \biggr] + \\
+ &\biggl[ \frac{1}{\Gamma_k+\tau\lambda} - \frac{1}{\Gamma_j+\tau\lambda} \biggr]\, \biggl[ \frac{\tau\lambda \bigl(\Gamma_k - \Gamma_j\bigr)}{\bigl(\Gamma_k+\tau\lambda \bigr) \bigl(\Gamma_j+\tau\lambda \bigr)} \biggr].
\end{split}
\end{equation}
Let us now consider the first line on the right hand side of \eq{eq:app2.14}. If $\Gamma_i > \Gamma_j$, the term in the first square brackets will be negative, while the term in the second square brackets is positive, and vice versa if $\Gamma_i < \Gamma_j$. Thus, the first line in \eq{eq:app2.14} is always non-positive. The same argument applies to the second and third line of \eq{eq:app2.14}, and thus the entire expression \eq{eq:app2.14} is non-positive. This shows that \eq{eq:app2.13} is non-positive and therefore $\partial_\lambda Q_{\measuredangle}(\lambda,\tau) \leq 0$.

\subsection{Convergence in system size}
In Sec.~\ref{sec:LGTs} of the main text, we constructed the effective Hamiltonian $\hat{\mathcal{H}}$ from the time-evolved operators $\hat{N}_0(t) = \sum_i \hat{n}_i(t)$, evaluated numerically in small systems. Here, we show that due to the locality of the relevant operators $\hat{n}_i(t)$, our approach generalizes to larger systems and converges in the thermodynamic limit.
For this purpose, we consider a two-dimensional lattice with a subsystem $A$ centered around the origin, whose linear length we denote as $r_A \sim \sqrt{|A|}$. Moreover, we introduce the operator $\mathbb{P}^{}_A$ that projects onto Pauli strings that are fully supported on $A$~\cite{tran2021_lr}.
Due to the locality of the PXP Hamiltonian $\hat{H}_0$, a time-evolved local operator $\hat{n}_i(t\leq \tau)$ originating at a site $i$ far from $A$ has exponentially small support on $A$. Formally, the standard Lieb-Robinson bound implies~\cite{lieb1972finite,tran2021_lr}  
\begin{equation} \label{eq:lr_bound1}
\| \mathbb{P}^{}_A \hat{n}_i(t\leq \tau) \| \lesssim e^{c ( v^{}_{\mathrm{LR}}\tau - (|i| - r_A)  )},
\end{equation}
where $v^{}_{\mathrm{LR}}$ is a finite Lieb-Robinson velocity and $c>0$. 
We now consider the projection $\mathbb{P}^{}_A \hat{N}_0(t)$ of the time-evolved \textit{global} number operator onto $A$. Due to \eq{eq:lr_bound1}, the combined contribution of operators $\hat{n}_i(t)$ originating at distances $|i|-r_A \geq L$ from the region $A$ is bounded (up to constant factors) by
\begin{equation}\label{eq:lr_bound2}
\begin{split}
\| \mathbb{P}^{}_A \sum_{i:\, |i|-r_A \geq L}\hat{n}_i(t\leq \tau) \|  &\lesssim \int_L^\infty dr\, r \, e^{c (v^{}_{\mathrm{LR}}\tau - r)} = \\
&= \frac{1+cL}{c^2} e^{c(v^{}_{\mathrm{LR}}\tau - L)}.
\end{split}
\end{equation}
As a consequence, $\mathbb{P}^{}_A \hat{N}_0(t)$ and the projection $\mathbb{P}^{}_A \hat{\mathcal{H}}[\Delta_p]$ of the effective Hamiltonian onto $A$ converge exponentially in system size when $L\gtrsim v^{}_{\mathrm{LR}}\tau$.
On the one hand, this implies that the effective Hamiltonian induced by a fixed detuning profile $\Delta_p(t)$ generalizes well to larger systems.
On the other hand, optimizing the projected $\mathbb{P}^{}_A \hat{\mathcal{H}}$ for large systems by minimizing the cost function
\begin{equation} \label{eq:lr_bound3}
C \equiv \| \mathbb{P}_A \hat{T} - \mathbb{P}_A \hat{\mathcal{H}}[\Delta_p] \|^2 + \lambda \|\Delta_p\|^2
\end{equation}
results in an exponentially converging optimized detuning profile $\tilde{\Delta}_p(t)$.

\end{appendix}
\newpage

%\bibliographystyle{plain}
\bibliography{floquet}

\end{document}